\documentclass{rspublic}
\usepackage{natbib}
\usepackage{graphicx}
\usepackage{dcolumn}
\usepackage{bm}
\usepackage{hyperref}
\renewcommand{\vec}[1]{{\mathbf{#1}}}
\newcommand{\beq}{\begin{eqnarray}}
\newcommand{\eeq}{\end{eqnarray}}
\renewcommand{\cite}{\citep}

\begin{document}

\title[Mottness Collapse and T-linear Resistivity]{Mottness Collapse and
  T-linear Resistivity in Cuprate Superconductors}

\author[P. Phillips ]{Philip Phillips}

\affiliation{Department of Physics,
University of Illinois
1110 W. Green Street, Urbana, IL 61801, U.S.A.}

\label{firstpage}

\maketitle

\begin{abstract}{Mottness, superconductivity, strange metal, pseudogap}
Central to the normal state of cuprate high-temperature
superconductors is the collapse of the pseudogap, briefly reviewed here, at a critical point and the subsequent onset of the
strange-metal characterized by a resistivity that scales linearly with temperature.  A possible clue to the resolution of this problem is
the inter-relation between two facts: 1) A
robust theory of T-linear resistivity resulting from quantum
criticality requires an additional length scale outside the standard 1-parameter scaling
scenario and 2) breaking the Landau correspondence between the Fermi gas
and an interacting system with short-range repulsions requires
non-fermionic degrees.  We show that a low-energy theory of the Hubbard
model which correctly incorporates dynamical spectral weight transfer
has the extra degrees of freedom needed to describe this physics. The degrees of
freedom that mix into the lower band as a result of dynamical spectral
weight transfer are shown to either decouple beyond a critical doping,
thereby signaling Mottness collapse or unbind above a critical
temperature yielding strange metal behaviour characterised by
$T-$linear resistivity. 
\end{abstract}

\section{Introduction}
High-temperature superconductivity in the copper-oxide ceramics
remains an unsolved problem because we do not know what the
propagating degrees of freedom are in the normal state. Consequently,
we cannot say with any certainty what are the weakly interacting
degrees of freedom that pair up to form the
superconducting condensate.  In low-temperature superconductivity in
metals, the existence of a Fermi surface simplified the identification
of the propagating degrees of freedom.   As shown by
Polchinski\cite{polchinski}, Shankar\cite{shankar} and others\cite{others}, all renormalizations arising
from short-range repulsive interactions are towards the Fermi
surface.  As a result,
such interactions can effectively be integrated out leaving
behind dressed electrons or quasiparticles as the propagating degrees
of freedom.  

Undoped, the cuprates are Mott insulators.  Charge localization
obtains in Mott systems, not because the band is full, and in fact it
is not, but because strong local electron correlations dynamically
generate a charge gap by splintering the half-filled band in
two. Hence, unlike low-temperature superconductors which are metals in
which the (short-range repulsive) interactions are irrelevant, Mott insulation obtains from
strong coupling physics. Precisely how the strong correlations mediate
the myriad of phases in the doped cuprates remains unsettled.
Nonetheless, there are some experimental facts which are clear. For example, when Mott insulators are doped, a `gap'
still remains\cite{pg} in the normal state.  Dubbed the pseudogap, as
zero-energy states exist\cite{norman} at some momenta, in particular along the
diagonal connecting $(0,0)$ and $(\pi,\pi)$, this phenomenon remains
one of the most nettling problems in cuprate phenomenology as many
of the articles in this volume attest. A typical value for the maximum
of the gap
along the $(\pi,0)$ direction in underdoped
Bi$_2$Sr$_2$CaCu$_2$O$_{8+\delta}$ (Bi2212) is approximately 45meV at
 an estimated hole content of $0.1$\cite{pg2gaps}. This
hole-content level was obtained, not by counting the number of dopant
atoms, as in the case of 
La$_{1-x}$Sr$_x$CuO$_2$ (LSCO), in which the number of doped holes is obtained
by counting the number of strontium atoms, but through the empirical formula\cite{preslund} 
\beq\label{empirical}
1-\frac{T_c}{T_c^{\rm max}}=82.6(x-0.16)^2 
\eeq
which accurately describes the evolution of the superconducting transition
temperature, T$_c$, for LSCO as a function of doping.  While this
formula is on firm experimental footing for LSCO, it has been
widely criticised in the context of cuprates such as
YBa$_2$Cu$_3$O$_{7-\delta}$ (Y123) and Tl$_2$Ba$_2$CuO$_{6+\delta}$
(Tl2201) in which it is the oxygen content that determines the doping
level.  For example, Tokura and colleagues\cite{tokura} and Tutsch, et
al\cite{tutsch} find optimal doping values of 
$x=0.21$ and $x=0.24$, respectively for YBCO.
Others\cite{merz,drechsler,markiewicz} have reached
similar conclusions.  Hardy and collaborators\cite{hardy} have
investigated the validity of Eq. (\ref{empirical}) for Y123.  However,
their work does not offer an independent check on the validity of
Eq. (\ref{empirical}) for Y123 because in correlating the change
in the c-axis lattice constant (see caption to Fig. 2\cite{hardy}) with the doping level, they used
Eq. (\ref{empirical}).  Consequently, they must confirm that
$p_{\rm opt}=0.16$ as they report.
\begin{figure}
\centering
\includegraphics[height=10.0cm]{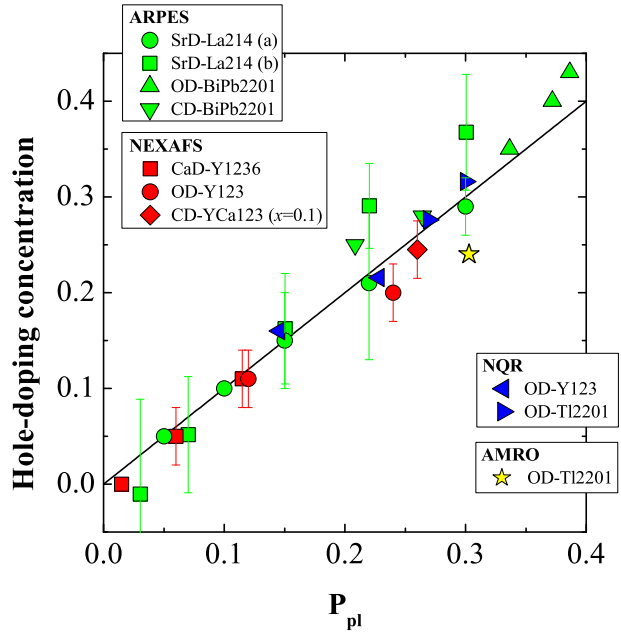}
\caption{Hole-doping level from various techniques compared with the
  doping scale extracted from the thermopower, $P_{\rm pl}$. The red
  points are obtained from (near-edge X-ray absorption fine structure)
  NEXAFS: red
  circles are OD-Y123\cite{ody123}, red diamonds are co-doped
  Ca-YaC123 (at x=0.1)\cite{ody123}, red squares are Calcium-doped
  Y1236\cite{ody123}.  The blue points are from nuclear-quadrupole
  resonance (NQR) measurements: left blue arrow is
  OD-Y123\cite{NQR} and the right blue arrow is
  OD-Ti2201\cite{NQR}.  The green points are from ARPES: green
  circles are Strontium-doped La214\cite{srd214a}, green squares are
  also Strontium-doped La214\cite{srd214b}, up arrow, overdoped
  BiPb2201\cite{arp1}, down arrow co-doped
  BiP2201\cite{arp1}.  The star corresponds to angular
  magnetoresistance oscillations (AMRO)\cite{amro}. Reprinted
from Phys. Rev. B {\bf 77}, 84520 (2008).}
\label{honmafig}
\end{figure} 

\begin{figure}
\centering
\includegraphics[width=7.0cm]{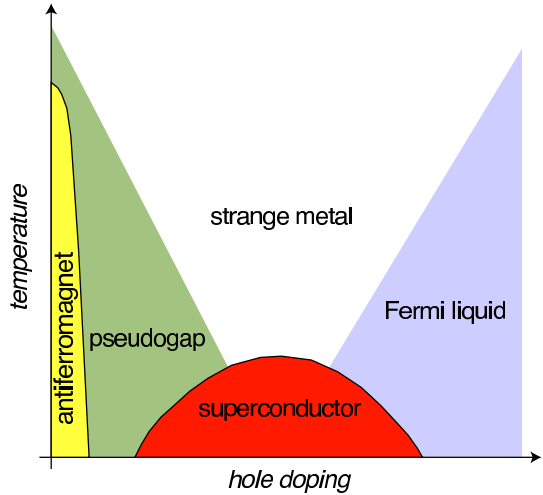}
\caption{Heuristic phase diagram of the copper-oxide
  superconductors.  In the
 strange metal, the resistivity is a linear function of
 temperature. In the pseudogap the single-particle density of states
 is suppressed without the onset of global phase coherence indicative
 of superconductivity.  As discussed in the introduction, the dome-shape of the superconducting region
 with an optimal doping level of $x_{\rm opt}\approx 0.17$ is not quantitatively
accurate any cuprate other than LSCO. }
\label{pd}
\end{figure}

I have digressed here on the doping
level in the cuprates because one of the key issues with the pseudogap
is where precisely it terminates as articulated in the review by
Norman, Pines and Kallin\cite{npk}.  If the superconducting regions of
all the cuprates are
artificially made to have optimal doping at a hole content of
$0.16$, then the pseudogap, as determined\cite{pg2gaps} by spectroscopic probes such
as angle-resolved photoemission (ARPES), scanning tunneling
microscopy, and Raman scattering, will terminate at the end of the
dome. While not all probes\cite{ando,tailifer} find that $T^*(x)$ merges with the terminus
of the superconducting dome, the problem is partially one of
accurately determining the doping level in the cuprates.  Luckily, Honma and Hor\cite{honma} have recently
addressed the problem of how to determine the doping level
unambiguously in the cuprates in which it is the oxygens that
act as the dopants.  They have advocated\cite{honma} that
because the room temperature thermopower data for {\it all} the cuprates 
collapses onto a single curve, that curve can be used to calibrate the
doping level and hence map out the superconducting region as a
function of doping in an unbiased fashion. I reprint here a figure
(see Fig. (\ref{honmafig}))
from their paper which illustrates that the thermopower scale
 is in excellent agreement with the doping level in Y123
determined by three different experimental methods.  With this scale,
the optimal doping level for most of the cuprates, except LSCO, occurs at $p_{\rm
  opt}\approx 0.21-0.23$\cite{honma}.  This is roughly the doping level at which
the pseudogap closes\cite{tailifer} as determined by
transport measurements\cite{ando}.   In this article, our focus is primarily
transport and hence we consider a phase diagram in which the
pseudogap crosses the superconducting dome as shown in
Fig. (\ref{pd}).  That the superconducting region is depicted with a
dome should not be taken literally. 

Several questions arise with the pseudogap.  Is it simply a remnant of
the Mott gap? Is it a precursor to superconductivity? Does it
represent a new ordered state? Does it compete with superconductivity?
 Regardless, of how one answers these questions, the phase diagram
 makes it abundantly clear that the correct theory of the pseudogap
 should above a characteristic scale, $T^*$, explain the strange metal
 in which the resistivity scales linearly with temperature.  It is
 indeed odd then that the
 plethora of models\cite{stripes1,ddw,rvb,inco1,inco2,inco3} proposed to explain pseudogap behaviour have had
 little to say about the strange metal as a robust phenomenon
 persisting to high temperatures.   Part of the problem is that a series of
associated phenomena, for example, incipient diamagnetism\cite{nernst}
indicative of
incoherent pairing\cite{inco1,inco2,inco3}, electronic
inhomogeneity\cite{stripes1,stripes2,stripes3,stripes4,stripes5,stripes6}, time-reversal
symmetry breaking\cite{trsb1,trsb2,trsb3,trsb4}, and quantum oscillations\cite{qoscill} in
the Hall conductivity,
possibly associated with the emergence of closed electron (not
hole) pockets in the first Brillouin zone (FBZ), are all consistent
with some type of order with no immediate connection with the strange
metal.   We review briefly some of the proposals for the pseudogap,
identify the problems in constructing a theory of the strange metal
and sketch a possible resolution of this problem based on our recent
work involving the exact integration of the high-energy scale in the
Hubbard model\cite{charge2e1,charge2e2,charge2e3,charge2e4,repprogphys}.

\section{Pseudogap Phenomena}

The phenomena surrounding the pseudogap\cite{pg,timusk,norman} in the cuprates used to be
fairly simple. In zero magnetic field, lightly doped cuprates possess Fermi arcs\cite{norman} in
the normal state.  That is, the Fermi surface which is present in the
overdoped, more conventional Fermi liquid regime is destroyed on
underdoping leaving behind only a Fermi arc.  In actuality, the
situation is much worse.  That the arc does not
represent a collection of  well-defined quasiparticles has been
clarified by Kanigel, et al.\cite{kanigel} who showed that in
Bi$_2$Sr$_2$CaCu$_2$O$_{8-\delta}$, the length of the Fermi arc
shrinks to zero as $T/T^*$ tends to zero.  Consequently, the only remnant of the
arc at $T=0$ is a quasiparticle in the vicinity of
$(\pi/2,\pi/2)$ and hence the consistency with nodal metal
phenomenology\cite{nodalmetal}.  Aside from ARPES\cite{kanigel}, Raman
data\cite{0603392} also support the nodal/antinodal dichotomy in the underdoped regime. 

Recently, however, new ingredients have been added to the pseudogap
story in the underdoped regime which, on the surface, are difficult to
reconcile with Fermi arcs.  At high magnetic fields, quantum oscillations, indicative of a
closed Fermi surface, have been observed\cite{tailifer} in Y123 and Tl-2201 through
measurements of the Hall resistivity, Shubnikov-de Haas effect, and
the magnetization in a de
Haas-van Alphen experiment.  In underdoped samples, the Hall
coefficient is negative indicating closed electron-like orbits.
Hence, there is an obvious incompatibility, though some have advocated
none occurs\cite{tailifer}, in interpreting the quantum
oscillation experiments as a zero-field property of the underdoped
cuprates where Fermi arcs obtain.
Further, the fields, roughly 50T, at which the quantum oscillations
are observed are insufficient to kill the large
gap that exists at the antinodal regions.  The recent proposal by Pereg-Barnea, et al.\cite{barnea} which shows that quantum oscillations can arise from Fermi arcs terminated by a pairing gap is noteworthy since it points to a possible resolution of the two phenomena.

Also attracting much attention is the recent experimental evidence for nematic
order\cite{bonnnematic,keimernematic}.  Daou, et al.\cite{bonnnematic}
observed that the Nernst signal in Y123 exhibits a large in-plane
anisotropy that increases with decreasing temperature.  The onset
temperature is roughly $T/T^*\approx 0.8$.  While the Nernst signal\cite{nernst}
had been measured previously\cite{nernst}, interpreted as evidence for
incoherent pair formation above $T_c$ but not all the way to $T^*$, it had not been measured with temperature gradients parallel
to the $a$ and $b$ axes separately.  While nematic
order\cite{stripes1} can give rise to an anisotropy in the
Nernst signal, Y123, in the doping regime measured, is already
anisotropic in the a-b plane as it is orthorhombic. Further, it has
been known for quite some time\cite{salamon} that the thermopower has
different signs parallel and perpendicular to the chains in Y123.    As
the Nernst signal involves the thermopower, such a sign difference could
drastically amplify any modest anisotropy that exists among the combination
of transport coefficients that contribute to the Nernst
signal. Consequently, further experiments are needed to disentangle
the inherent $a$/$b$ axis asymmetry in Y123 from that arising from an
electronic phase that spontaneously breaks rotational symmetry.     

Theoretical proposals for the pseudogap fall into three groups: 1)
Mott-related physics having nothing to do with order\cite{kotliar,rvb,mottness}, 2) precursor
superconductivity\cite{inco1,inco2,inco3,senthil}, and 3) ordered states that break some type of
symmetry, be it 4-fold
rotational symmetry\cite{stripes1}, translational
symmetry\cite{stripes2,ddw}, or time-reversal
symmetry\cite{trsb3}. In fact, experimental data support\cite{bonnnematic,trsb1,trsb2,keimernematic}
many of the ordered states proposed for the onset of the pseudogap. 
However, it is unclear how any of these proposals are related to the
origin of the strange metal. Nonetheless, a common approach\cite{pchamon} to meld ordered states
with $T-$linear resistivity is to invoke quantum criticality. However,
in its simplest one-parameter form, quantum criticality fails\cite{pchamon} to yield
$T-$linear resistivity unless the dynamical critical exponent is
negative, thereby violating causality.  This result follows from three
simple assumptions: 1) the charges are critical, 2) one-parameter
scaling is valid, and 3) the $U(1)$ charge is conserved.  These three assumptions
yield immediately to a general scaling form for the conductivity
\begin{eqnarray}\label{dclimit}
\sigma(\omega=0)=\frac{Q^2}{\hbar}\;\Sigma(0)\;\left(\frac{k_BT}{\hbar
   c}\right)^{(d-2)/z} 
\end{eqnarray}
with $Q$ the charge, $\Sigma(0)$ the conductivity at $\omega/T=0$, and $z$ is the dynamical
exponent.  
As a result, quantum criticality in its present form yields
$T-$linear resistivity (for d=3) only if the dynamical exponent
satisfies the unphysical constraint $z<0$.  The remedy here might be
three-fold:  1)  some other yet-unknown phenomenon  is responsible for
$T-$linear resistivity, 2) the charge carriers are non-critical, or 3)
the single-parameter scaling hypothesis must be relaxed. 

Marginal
Fermi liquid theory\cite{mfl} (MFL) does, however, offer a phenomenological account of
$T-$linear resistivity. The key posit\cite{mfl} here is that at $T=0$, regardless of
momentum, the single-particle
scattering rate is proportional to $|\omega|$. Because of the linear
dependence of the scattering rate on frequency, MFL phenomenology
yields immediately $T-$linear resistivity.  However, at present, there
is no microscopic derivation of this highly successful
account.  Recently, however, the gauge/gravity duality\cite{maldacena} has proved useful\cite{lmv}
in this context.  The key claim of the gauge/gravity duality is that
some interacting quantum theories at strong coupling in d-space-time
dimensions are dual to gravity theories in $d+1$ dimensional
asymptotically anti-de Sitter (AdS$_{d+1}$) space-time.  Unlike the
standard 
equivalence between
partition functions for
d-dimensional quantum and $d+1$-dimensional classical systems in which
the extra dimension represents time, in the gauge/gravity duality, the
extra dimension represents the renormalization group (RG) scale. That is, in the AdS$_{d+1}$
construction, an infinite number of copies of the original quantum mechanical theory, each at a
different RG scale, fill the extra dimension.  Hence, the original
strongly coupled theory lives entirely at the boundary of the
AdS$_{d+1}$ space.  In this construction, finite temperature and
finite density correspond to having a charged black hole in the bulk
geometry.  Since the gravity theory is purely classical, all questions
surrounding the strong-coupling physics in the original problem can be
obtained from solving a set of linear wave equations in the charged black-hole
geometry.   Using the AdS$_4$ construction, Liu, McGreevy, and Vegh\cite{lmv}
 computed the single-particle electron spectral function and
showed that a range of non-Fermi liquid self-energies can emerge including
that of marginal Fermi liquid theory. This result is truly remarkable
as it represents the only derivation, albeit not microscopic, of marginal Fermi liquid
theory.  However, a key question remains.  Liu, Mcgreevy and
Vegh\cite{lmv} worked entirely with the gravity theory, and hence the
underlying quantum theory is not known.  Nonetheless, some hints as to the
nature of the underlying theory are contained in the gravity solution
to marginal Fermi liquid theory.  A key feature of the quantum
theory on the charged AdS$_4$ geometry is that the IR or low-frequency
analytic structure of the correlation
functions for the charge degrees of freedom are determined entirely by
the near horizon metric AdS$_2$ $\times \mathbb R^2$. This serves to
illustrate that the degrees of freedom which emerge in the IR and
govern the analytic structure of the theory have no correspondence to
those in
the original UV charged AdS$_4$ limit.  Perhaps the same is true of
the underlying microscopic quantum theory which ultimately describes
marginal Fermi liquid theory.  This suggests that extracting
marginal Fermi liquid behaviour from the basic model for a doped Mott
insulator might be tricky because the natural variables which would
expose this behaviour are not the bare electrons.  Within the Hubbard
model, it is dynamical spectral weight transfer across the Mott gap that makes the construction of a low-energy
theory difficult. Isolating the propagating degrees of freedom which
make the low-energy physics weakly interacting amounts to choosing a
set of variables which essentially gets rid of the dynamical mixing.
Such variables should in principle hold the key to $T$-linear
resistivity.   It is precisely this problem that we now address.

\section{Fermi Liquid Theory Breakdown: Dynamical Spectral Weight Transfer}

To isolate how dynamical spectral weight transfer leads
to a breakdown of Fermi liquid theory,
it suffices to focus on how the intensity of the band in which the
chemical potential resides scales with doping. For hole doping, the
relevant band is the lower Hubbard band (LHB). The intensity of a band
is equal to the number of charges that can
fit into that band.   A problem arises for
a Fermi liquid account whenever there is a mismatch between the
intensity of a band and the number of electrons that can fill the
band.  Such is the case in the Hubbard model close to half-filling.  
In the atomic limit, everything is known exactly.  There are two bands
split by the on-site repulsion $U$.  When $x$ holes are introduced, the intensity of the
LHB is $1+x$ and that of the upper Hubbard band (UHB) is $1-x$.  The weight
in the LHB separates into two parts.  Since each hole leaves an empty
site that can be occupied by either a spin-up or a spin-down electron,
the empty part of the LHB has weight $2x$\cite{sawatzky} 
and the occupied part an
intensity of $1-x$.  Hence, we see that the weight of the UHB and the
occupied part of the LHB are equal in the atomic limit.  This follows
from the simple fact that removal of a doubly occupied site also
removes one state from the LHB as well.  There are, however, $t/U$ corrections to these intensities
beyond the atomic limit.  In 1967, Harris and Lange\cite{hl} showed
that the intensity of the LHB
 \beq\label{hleq}
{\rm m}_{\rm LHB}
	=1+x+\frac{2t}{U}\sum_{ij\sigma}g_{ij}
		\langle f_{i\sigma}^\dagger f_{j\sigma}\rangle+\cdots
	=1+x+\alpha,
\eeq
has $t/U$ corrections\cite{hl} which are entirely positive.  Here 
  $f_{i\sigma}$ is related to the original bare fermion operators via a 
canonical transformation that brings the Hubbard model into block diagonal 
form in which the energy of each block is $nU$. In fact, all orders of 
perturbation theory\cite{eskes,hl} increase the intensity of the LHB beyond its 
atomic limit of $1+x$. It is these dynamical corrections that $\alpha$ 
denotes. While the intensity of the LHB increases away from the atomic limit, 
the total number of ways of assigning electrons to the LHB still remains fixed 
at $1+x$. That is, the number of electron states in the LHB is indepenent of the hopping. Consequently, there is a mismatch between the intensity of
the LHB and the number of electrons this band can hold.  Since $m_{\rm
  LHB }>1+x$, additional non-fermionic degrees of freedom are needed
to exhaust the phase space of the LHB.  These degrees of freedom
affect the physics at all energy scales in the LHB.   This has a
profound consequence.  The conserved charge is still the electron
filling, $n_e$ which is obtained by integrating the density of states
in the LHB up to the chemical potential, appropriately defined.  But
this quantity now has two contributions,
\beq\label{ne}
n_e=n_{\rm qp}+n_{\rm nf},
\eeq
 one
coming from the fermionic low-energy degrees of freedom, $n_{\rm qp}$, in the LHB and the
non-fermionic part.  As a result, the fermionic quasiparticles in the
LHB and the
bare electrons can no longer correspond one-to-one as $n_e>n_{\rm qp}$.  This constitutes
a breakdown of Fermi liquid theory. 

\begin{figure}
\includegraphics[width = 8.0cm]{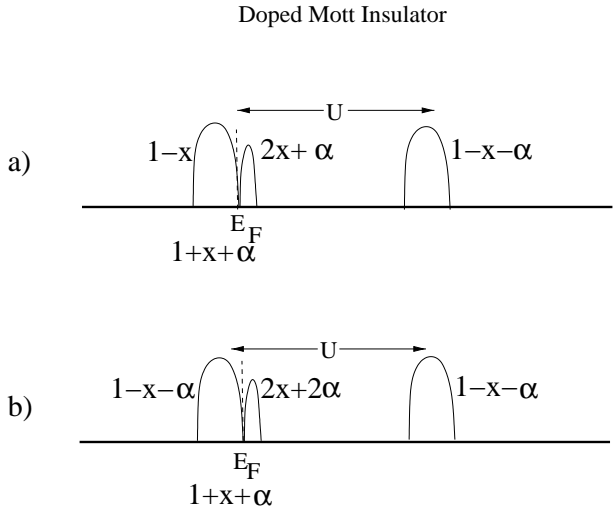}
\caption{Redistribution of spectral weight in the Hubbard model upon doping the
		insulating state with $x$ holes. $\alpha$ is the dynamical correction 
		mediated by the doubly occupied sector. To order $t/U$, this correction 
		worked out by Harris and Lange\cite{hl}. a) The traditional approach
		\cite{sawatzky,stechel} in which the occupied part of the 
		lower band is fixed to the electron filling $1-x$.  b)
                New assignment of the spectral weight in terms of
                dynamically generated charge carriers.  In this
                picture, the weight of the empty part of the LHB per spin is
                the effective doping level, $x'=x+\alpha$. }
\label{spec}
\end{figure}

We propose that the chemical potential for the effective number
of low-energy fermionic
degrees of freedom can be determined by partitioning the
spectrum in the LHB so that dynamical spectral weight transfer is
essentially removed.  In such a picture, the empty part of the
spectrum per spin is equal to the weight removed from the occupied
part of the LHB when a hole is created.
 Hence, we arrive at the assignments of the spectral weights in
Fig. (\ref{spec}b) in which the doping level is renormalized by the
dynamics; that is, $x'=x+\alpha$.  That
is, the effective number of fermionic degrees of freedom in the LHB is  
less than the conserved charge. This result already follows from the
fact that if the intensity of the LHB exceeds $1+x$, electrons alone
cannot exhaust the total degrees of freedom at low energies.  Hence,
by Eq. (\ref{ne}), there are two contributions to the conserved charge,
thereby implying that the number of fermionic quasiparticles is less than the
conserved charge.  In other words, the 
dynamical degrees of freedom denoted by $\alpha$ serve to supplement the effective 
phase space of a hole-doped system and $x'=x+\alpha$ now denotes the effective 
number of hole degrees of freedom per spin at low energy. 

\begin{figure}
\centering
\includegraphics[width=8.0cm]{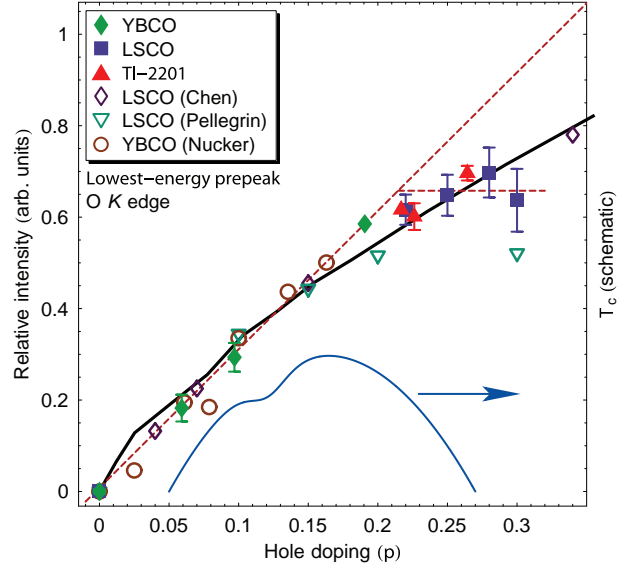}
\caption{ Reprinted from Phys. Rev. Lett. {\bf 103},
  087402 (2009). Compilation of the doping dependence of the lowest-energy oxygen K-edge
  pre-peak from four data sources. The solid plotting symbols are from
  Peets, et al.,\cite{peets}. The open diamonds are taken from Chen, et
  al.\cite{chen}, open triangles from Pellegrin, et
  al.\cite{pellegrin} and open circles from N\"ucker, et
  al.\cite{nucker}. The solid straight line is the low-energy spectrum in the Hubbard model computed by A. Liebsch\cite{liebsch}. A constant scale factor was used to collapse the Liebsch points onto the experimental data since the units of experimental data are arbitrary.   The superconducting dome is indicated for
  reference. }
\label{mottcollapse}
\end{figure}

For sufficiently large doping levels, the UHB collapses and the
standard weakly interacting picture emerges. At this point, it is no
longer meaningful to expand around the atomic limit and the analysis
leading to Fig. (\ref{spec}b) fails.   This failure arises because,
the non-interacting ground state is not
adiabatically connected to the atomic limit.  Rather, perturbation
theory around the band limit should be performed.  In this limit, the
weight of the band in which the chemical potential resides is $2$,
completely independent of doping.   Hence, there must be a critical point as a
function of doping or interaction strength at which the intensity of the lower band jumps to
$2$.  This constitutes a collapse of Mottness.  We have advocated\cite{repprogphys} that
the doping level at which the UHB collapses coincides  with the
closing of the pseudogap.  This is physically reasonable because
unless a gap exists, there is no real separation between the UHB and
LHB. The simulations of Kyung, et al.\cite{kyung} on the Hubbard model
also provide clear evidence that a strong correlation exists between
the pseudogap and the separation of the low and high energy bands.  Experimental evidence for this collapse has been reported
recently by Peets, et al.\cite{peets} from soft x-ray scattering on
the oxygen K-edge.  In such an experiment, an electron is promoted from the core 1s to an
unoccupied level. The experimental observable is the fluorescence
yield as a function of energy as electrons relax back to the
valence states.  Experimentally\cite{peets}, the fluorescence yield is related to the empty part of the spectrum projected onto the oxygen p-orbitals. Consequently, within a 1-band Hubbard model, the relevant quantity is  
\beq
L=\int_\mu^\Lambda N(\omega)d\omega.
\eeq
As shown in Fig. \ (\ref{mottcollapse}),
beyond a critical doping level, the slope of the oxygen K-edge intensity changes abruptly.  Peets, et al.\cite{peets} interpreted the slope change as evidence for a saturation and hence a deviation from what is expected in the 1-band Hubbard model.  However, given the error bars on the data and the additional point at $x=0.34$ of Chen, et al.\cite{chen}, one cannot rule out that the data are simply associated with a slope change around a doping level of $x=0.22$. To verify this assertion, we plotted (Fig. \ (\ref{mottcollapse})) a computation of $L$ (solid line) by A. Liebsch\cite{liebsch} on the 2-d Hubbard model using a self-consistent cluster method.  As is evident, the agreement with the solid curve and most of the data points is excellent.  Hence, over the complete doping range of interest,  the low-energy spectral weight in the cuprates is well described by the doping-induced states in the 1-band Hubbard model.

To summarize,  dynamical spectral weight has two profound effects in
the normal state of a doped Mott insulator.  As long as it is present,
a low-energy theory with electrons alone cannot exhaust the total
number of charge degrees of freedom in the LHB. This results in a mismatch
between the number of fermionic quasiparticles and bare electrons,
thereby leading to a breakdown of Fermi liquid theory.  In addition,
$L/n_h>1$ implies that the number of ways of adding particles at low energies exceeds
the number of electrons that can be added to the LHB. Consequently, $L/n_h>1$ is
an indication that some states at low energy are gapped to the
addition of an electron.  Within the experimental accuracy\cite{peets},
there is a dramatic slope change 
from 
in $L$ in the vicinity of where the pseudogap closes, adding
credence to the emphasis
placed here on dynamical spectral weight transfer as the signature of
the pseudogap.  However, to be
completely consistent with the opening of a pseudogap, dynamical
spectral weight transfer should also be temperature dependent.  We show here
that this is the case.  Namely, the dynamical correction turns on only
below a characteristic temperature set by $T^\ast$. This is an
indication that some type of bound state formation underlies
$\alpha\ne 0$.  Above $T^\ast$,
the new degrees of freedom are unbound and create the $T-$linear
resistivity.  Since this mechanism is capable of generating $T-$linear
resistivity beyond $T^\ast$ and a pseudogap below, we arrive at a
consistent theory of two of the most elusive characteristics of the
normal state of the cuprates.

\section{Low-energy Theory of  Mottness}

In the parameter space relevant to the cuprates, the Hubbard model represents strongly coupled physics.  In this regime, no continuum limit exists and as a consequence, no firm statement can be made as to the precise nature of the charge carriers. Ultimately, it is only correct in this regime to focus on the current, which can be given an interpretation in terms of single-particles if the propagating degrees of freedom can be identified.  Such an identification should be possible once the high-energy scale is correctly integrated out.  A successful 
low-energy reduction of the Hubbard model should reveal
the new degree of freedom that causes the intensity of the lower band to
exceed $1+x$.  As this degree of freedom arises from the mixing
with the doubly occupied sites, it should have charge $2e$. Charge $2e$
objects do not contribute to the fluorescence yield of the single-particle spectrum unless they
combine with something else to produce a charge $e$ entity.  The proposed
mechanism for the pseudogap is `doublon-holon' bound-state
formation mediated by a collective IR charge $2e$ mode.   Below a
characteristic temperature, such bound states are stable. Quite
generally, such bound
states are expected to mediate the Mott insulating state as well.  Consider
half-filling.  Even in the half-filled state, the dynamical mixing
that leads to a non-zero $\alpha$ is still present.  That is, double
occupancy is present even in the ground state of a half-filled band.
Since a doubly occupied site must result in the simultaneous creation
of an empty site, the insulating state persists only if the doubly
occupied and empty site are bound as has been proposed
previously\cite{castellani,kohn1, mott}.  Else, hole conduction obtains.

\begin{figure}
\centering
\includegraphics[width=8.0cm,angle=0]{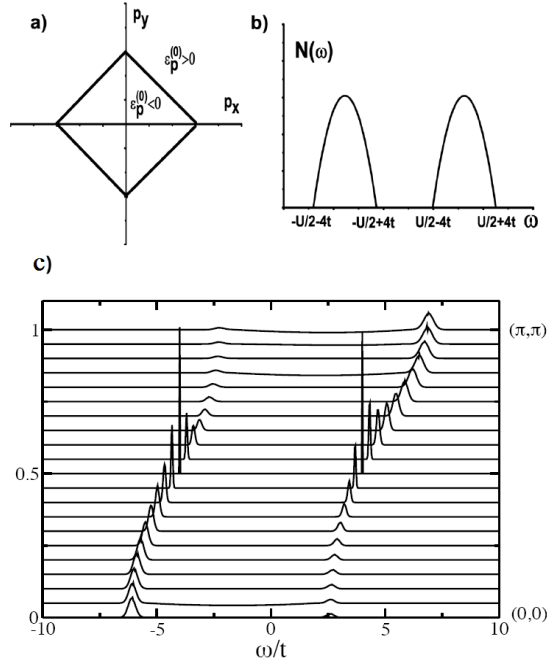}
\caption{a) Diamond-shaped surface in momentum space where the
  particle dispersion changes sign.  b) Turn-on of the spectral weight in the upper and lower Hubbard
  bands for the composite excitations as a function of energy and momentum.  In the UHB of the composite excitations, the
  spectral density is determined to $\gamma_{\vec p}$  while for the LHB it is
  governed by $\tilde\gamma_{\vec p}$.  The corresponding operators
which describe the turn-on of the spectral weight are the composite
excitations $\varphi^\dagger cc$ (UHB) and $\tilde\varphi cc$ (LHB).}
\label{mottgap}
\end{figure}

Such binding should emerge from the correct low-energy theory of a
half-filled Hubbard band.  
We have shown previously\cite{charge2e,charge2e1,charge2e2,charge2e3,charge2e4,mottness}, how the high-energy degrees of freedom can be
integrated out exactly at any filling.  The key idea is\cite{charge2e,charge2e2,charge2e3,charge2e4} to extend the
Hilbert space by introducing a new fermionic field that creates
excitations on the $U$ scale without identifying such physics with
double occupancy.  At half-filling the Lagrangian\cite{charge2e4}
simplifies to
\beq\label{huv}
L^{\rm hf}_{\rm IR}&=& 2\frac{|s|^2}{U}|\varphi_\omega|^2+2\frac{|\tilde s|^2}{U}|\tilde\varphi_{-\omega}|^2+ \frac{t^2}{U} |b_\omega|^2\label{eq:actlineone}\\
&&+s\gamma_{\vec p}^{(\vec k)}(\omega)\varphi^\dagger_{\omega,\vec
  k}c_{\vec k/2+\vec p,\omega/2+\omega',\uparrow}c_{\vec k/2-\vec
  p,\omega/2-\omega',\downarrow}\nonumber\\
&+&\tilde s^*\tilde\gamma_{\vec p}^{(\vec k)}(\omega)\tilde\varphi_{-\omega,\vec k}c_{\vec k/2+\vec p,\omega/2+\omega',\uparrow}c_{\vec k/2-\vec p,\omega/2-\omega',\downarrow}+h.c.,\label{eq:actlinetwo}.
\eeq 
This theory contains two bosonic fields with charge $2e$
($\varphi^\dagger$) and $-2e$ ($\tilde\varphi$). These bosonic modes are collective degrees of
freedom, not made out of the elemental excitations, which represent dynamical mixing with $U-$scale physics,
namely, the contribution of double holes ($-2e$) and
double occupancy ($2e$) to any state of the system.   Here
$s$ and $\tilde s$ are constants with units of energy, all
operators in Eq. (\ref{huv}) have the same site index, repeated indices
are summed both over the site index and frequency, $\omega$,
$c^\dagger_{i\sigma}$ creates a fermion on site $i$ with spin
$\sigma$, 
\beq\label{fb}
b_{\vec k}=\sum_{\vec p}\varepsilon^{(\vec k)}_{\vec p}\ c_{\vec k/2+\vec p,\uparrow}c_{\vec k/2-\vec p,\downarrow},
\eeq
and the dispersion is given by $\varepsilon^{(\vec k)}_{\vec p}=4\sum_\mu\cos(k_\mu a/2)\cos(p_\mu
a)$, where $\vec k$ and $\vec p$ are the center of mass and relative
momenta of the fermion pair.   The coefficients 
\beq
\gamma_{\vec p}^{(\vec k)}(\omega)&=&\frac{-U+t\varepsilon_{\vec p}^{(\vec k)}+2\omega}{U}\sqrt{1+2\omega/U}\nonumber\\
\tilde\gamma_{\vec p}^{(\vec k)}(\omega)&=&\frac{U+t\varepsilon_{\vec p}^{(\vec k)}+2\omega}{U}\sqrt{1-2\omega/U}
\eeq
play a special role in this theory as they account for the turn-on of
the spectral weight.  At the level of a Lagrangian, the vanishing of
the coefficient of a quadratic term defines the dispersion of the
associated particle.  All the terms which are naively quadratic,
Eq. (\ref{eq:actlineone}), possess constant coefficients and hence we
reach the conclusion that there are no propagating bosons or
electrons. What
 Eq. (\ref{eq:actlineone}) lays plain is that the turn-on of the spectral
 weight in a Mott insulator cannot be formulated in terms of bare
 electrons, at least in a low-energy theory. This is consistent with the emerging view\cite{zeros1,zeros2,zeros3} that in a
 Mott insulator, the single-particle Green function vanishes along
a locus of points in momentum space.  Physically, a vanishing of the
single-particle electron Green function implies that the electrons are
not the propagating degrees of freedom.  Precisely what the propagating
degrees of freedom are is determined by the dispersing modes in the
low-energy Lagrangian.  Consider the second line of the Lagrangian, Eq. (\ref{eq:actlinetwo}).  Appearing here are two
interaction terms, which describe composite excitations, whose coefficients can vanish.  Fig. (\ref{mottgap}b)
shows explicitly that the vanishing of $\gamma$ and $\tilde\gamma$
leads to spectral weight which is strongly peaked at two distinct
energies, $\pm U/2$.  Each state in momentum space has spectral weight
at these two energies.  The width of the bands is $8t$. The particles
which give rise to the turn-on of the spectral weight are composite
excitations or the bound states of the bosonic and fermionic degrees
of freedom determined by the interaction terms $\varphi^\dagger cc$
and $\tilde\varphi cc$.  In the terms of the variables appearing in the Hubbard model, we make the heuristic association of the composite excitations with bound states of double occupancy and holes as has been postulated previously\cite{castellani} to be the ultimate source of the gap in a Mott insulator.  This association is entirely heuristic because the variables the variables in the original UVIn so far as they generate the spectral
weight, the interaction terms can be thought of as the kinetic terms
in the low-energy action.  The gap (Mott gap) in the spectrum for the composite excitations obtains for $U>8t$ as each band is centered at $\pm U/2$ with a width of $8t$. Fig. (\ref{mottgap}) demonstrates that the transition to the Mott insulating state
found here proceeds by
a discontinuous vanishing of the spectral weight at the chemical
potential to zero but a continuous evolution of the Mott gap as is
seen in numerical calculations\cite{imada} in finite-dimensional lattices but not
in the $d=\infty$\cite{dinfty} solution.   

In terms of the bare electrons, the overlap with the composite
excitations determines the Mott gap.  To determine the overlap, it is tempting to complete the
square on the $\varphi^\dagger cc$ term bringing it into a quadratic form,
$\Psi^\dagger\Psi$ with $\Psi=A\varphi+B cc$.  This would lead to
composite excitations having charge $2e$, a vanishing of the overlap
and hence no electron spectral
density of any kind.  However, the actual excitations that
underlie the operator $\varphi^\dagger cc$ correspond to a
linear combination of charge $e$ objects, $c^\dagger$ and $\varphi^\dagger c$.  In terms of the UV variables, the latter can be
thought of as a doubly occupied site bound to a hole.  At half-filling\cite{charge2e1,charge2e3}, the exact representation of the electron creation
\beq\label{cop1}
c_{i,\sigma}^\dagger\rightarrow \tilde c_{i,\sigma}^\dagger&\equiv& -
V_\sigma\frac{t}{U}\left(c_{i,-\sigma}b_i^\dagger + b_i^\dagger c_{i,-\sigma}\right)\nonumber\\
&+&V_\sigma\frac{2}{U}\left(s \varphi_i^\dagger + \tilde s \tilde\varphi_i\right) c_{i,-\sigma}
\eeq
is indeed a sum of two composite excitations, the first having to do
with spin fluctuations ($b^\dagger c$) and the other with
high-energy physics, $\varphi^\dagger c$ and $\tilde \varphi c$, that
is, excitations in the UHB and LHB, respectively. 
We can think of the overlap 
\beq\label{ov}
O=|\langle c^\dagger|\tilde c^\dagger\rangle\langle \tilde
c^\dagger|\Psi^\dagger\rangle|^2 P_\Psi
\eeq
in terms of the physical process of
passing an electron through a Mott insulator.  The overlap will
involve that
between the bare electron with the low-energy excitations of Eq. (\ref{cop1}), $\langle c|\tilde
c\rangle$, and the overlap with the propagating degrees of freedom,
$\langle \tilde c|\Psi\rangle$ with $P_\Psi$, the 
propagator for the composite excitations.  Because of the
dependence on the bosonic fields in Eq. (\ref{cop1}), $O$ retains
destructive interference between states above and below the chemical
potential.  Such destructive
interference between excitations across the chemical potential leads to a
vanishing of the
spectral weight at low energies\cite{sawatzky}.  Consequently, the
turn-on of the {\it electron} spectral weight cannot be viewed simply as a sum of the spectral weight
for the composite excitations.  As a result of the destructive interference, the
gap in the electron spectrum will always exceed that for the composite
excitations.  Hence,
establishing (Fig. (\ref{mottgap})) that the composite excitations display a gap is a
sufficient condition for the existence of a charge gap in the electron
spectrum.  A simple calculation\cite{charge2e4}, Fig. (\ref{mottgap}c), of the electron spectral function at
$U=8t$ confirms this basic principle that a gap in the propagating
degrees of freedom guarantees that the electron spectrum is gapped.
Further, Fig. (\ref{mottgap}) confirms that the electron spectral
function involves interference across the Mott scale.  Consequently,
although the composite excitations are sharp, corresponding to
poles in a propagator as in Eq. (\ref{eq:actlinetwo}), the electrons are
not.

\subsection{Binding-Unbinding Transition}

The bound states that mediate the Mott gap also survive in the doped
state.  In fact, through dynamical spectral weight transfer, such
bound states are
transferred down in energy from the UHB. 
That some type of bound state must exist in the doped state of a
Mott insulator can be inferred from the work of Gor'kov and
Teitel'baum\cite{gorkov}.  They 
observed remarkably that the charge carrier concentration, n$_{\rm Hall}$,
extracted from the inverse of the Hall coefficient in
La$_{2-x}$Sr$_x$CuO$_4$ (LSCO) obeys an empirical formula,
\beq\label{n} 
n_{\rm Hall}(x,T)= n_{0}(x) + n_{1}(x)\exp(-\Delta(x)/T),
\eeq
appropriate for a two-component or two-fluid system.  One of the
components is independent of temperature, n$_0(x)$ ($x$ the doping level) while the other is
strongly temperature dependent, $n_{1}(x)\exp(-\Delta(x)/T)$.  
The key observation here is that the temperature dependence in
$n_{\rm Hall}$ is carried entirely within $\Delta(x,T)$ which defines a
characteristic activation energy scale for the system, the pseudogap
scale.  Hence, a charge density that obeys Eq. (\ref{n}) is
symptomatic of a gap in the spectrum.  Eq. (\ref{n}) offers direct
confirmation for a renormalization of the doping level purely from
dynamical effects. Our key contention here is that $\alpha$ arises
from the second term in Eq. (\ref{n}) and is mediated by the binding
of $\varphi_i^\dagger$ to a hole. 

\begin{figure}
\centering
\includegraphics[width=13.0cm]{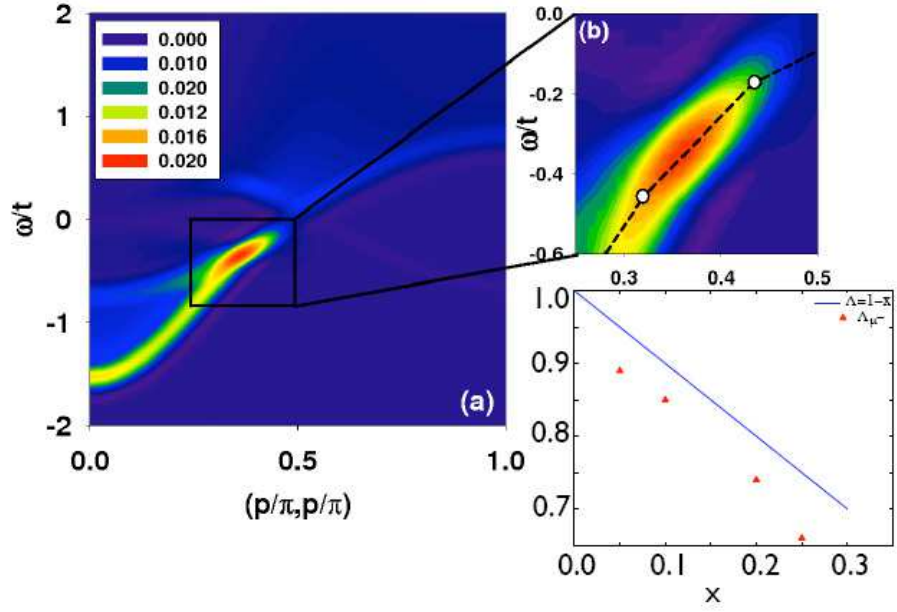}
\caption{(a) Spectral function for filling $n=0.9$ along the nodal
  direction.  The intensity is indicated by the color scheme.  (b)
  Location of the low and high energy kinks as indicated by the change
  in the slope of the electron dispersion.  (c)Integrated weight (triangles) of the fermionic part of the spectral function.  The deviation from $1-x$ (straight line) stems from the form of the conserved charge in Eq. (\ref{conscharge}).}\label{specf1}
\end{figure}

We computed the Hall coefficient\cite{hallcalc}
in this theory by first obtaining the spectral function.  As the
computation of the spectral function is instructive in uncloaking the
new propagating degrees of freedom, we recount it here. Noting that the conserved
charge is given by
\beq\label{conscharge}
Q=\sum_i c_{i\sigma}^\dagger
  c_{i\sigma}+2\sum_i\varphi^\dagger_i\varphi_i,
\eeq
the spectral function for just the fermionic component at low energies
$c_{i\sigma}$ will necessarily have an integrated weight less than
$1-x$\cite{dswtfinal1}.  To illustrate this, we treat $\varphi_i$ initially as a
spatially independent field, owing to the lack of any gradient terms
of $\varphi_i$ in the action.  Note, that the integrated weight of the
UHB is less than $1-x$ follows strictly from the form of the conserved
charge not from any approximation scheme that is used to compute the
spectral function.  
The electron Green function is then written as a path integral over the $\varphi$ fields as
\beq
\label{Green}
G(\bf{k},\omega)&=&\frac{1}{Z}
\int[D\varphi^*][D\varphi]FT(\int[dc^*_i][dc_i]c_i(t)c^*_i(0)\nonumber\\
&\times&\exp{\left(-\int L(c,\varphi) dt\right)})
\eeq
where the effective Lagrangian $L$ is expressed in a diagonalized form
\beq
L&=& \displaystyle \sum_{k\sigma}\gamma^*_{k
  \sigma}\dot{\gamma}_{k\sigma}+\displaystyle \sum_k (E_0 + E_k -
\lambda_k) + \displaystyle \sum_{k \sigma}
\lambda_k \gamma^*_k \gamma_k,\nonumber\\
\label{leff}
\eeq
where the $\gamma_{k\sigma}$ are the Boguliubov quasiparticles and are given by
\beq\label{bogoqp}
\gamma^*_{k\uparrow}= \cos\theta_k c^*_{k\uparrow} +\sin\theta_k c_{-k\downarrow}
\eeq
\beq\label{bogoqp2}
\gamma_{k\downarrow}=-\sin\theta_k c^*_{k\uparrow} + \cos\theta_k c_{-k\downarrow}
\eeq
where $\cos^2\theta_k =\frac{1}{2}(1 + \frac{E_k}{\lambda_k})$, $\alpha_k = 2(\cos k_x + \cos k_y)$ , $E_0 = (-2\mu + \frac{s^2}{U})\varphi^*\varphi
$,$E_k = -g_t t\alpha_k -\mu$,$\lambda_k = \sqrt{E^2_k
  +\Delta^2_k}$, the gap is proportional to $s$, $\Delta_k =
s\varphi^*(1-\frac{2t}{U}\alpha_k)$, and hence vanishes when $\varphi$
is absent and
$g_t = \frac{2\delta}{1 + \delta}$, $\delta = 1 -n$.  The $g_t$ term
originates from the correlated hopping term,
$(1-n_{i\bar\sigma})c_{i\sigma}^\dagger c_{j\sigma}(1-n_{j\bar\sigma})$.
The
The $\gamma_{k\sigma}$'s play the role of the fundamental low energy
degrees of freedom in a doped Mott insulator. That is, they are the
natural propagating charge degrees of freedom.  Note they depend in a
complicated way on the the $\varphi_i$ field and consequently are
heavily mixed with the doubly occupied sector.  Upon integrating out
the fermions, we obtain, 
\beq\label{eqG}
G(k,\omega) = \frac{1}{Z}\int [D\varphi^*] [D\varphi] G(k,\omega, \varphi) \exp^{-\sum_k (E_0 + E_k - \lambda_k -\frac{2}{\beta} \ln(1+e^{-\beta\lambda_k}) )}
\eeq
where
\beq\label{gfinal}
G(k,\omega,\varphi)=\frac{\sin^2\theta_k[\varphi]}{\omega+\lambda_k[\varphi]} + \frac{\cos^2\theta_k[\varphi]}{\omega-\lambda_k[\varphi]}
\eeq
is the exact Green function corresponding to the Lagrangian, Eq. (\ref{leff}),
which has a two-branch structure, corresponding to the bare electrons
and the coupled holon-doublon state respectively. The role of the
$\varphi$ field, which determines the weight of the second branch, is
vital to our understanding of the properties of Mott systems, as
was demonstrated previously\cite{charge2e1,charge2e3,repprogphys}. It is trivial  to see that in the
limit of vanishing $s$ (no $\varphi$ field), the $\gamma_{k\sigma}$'s
reduce to the bare electron operators $c_k$ and the first term in
Eq.\ (\ref{gfinal}) vanishes.   The two-fluid nature of the response
stems from this fact of the theory.  Namely, the first term
contributes only when $\varphi\ne 0$ and the second when $\varphi=0$
as depicted in the spectral function in Fig. (\ref{specf1}a).
These contributions correspond to the dynamical and static components of the
spectral weight, respectively and appear as two distinct branches in
the electron spectral function.  Since the second branch corresponds
to a bound degree of freedom, a gap opens at the chemical potential.
As we demonstrate in Fig. (\ref{specf1}c), the total weight of both branches which composes the total number of fermionic degrees of freedom is less than the conserved charge $1-x$.  This is dictated by the fact that conserved charge at low energies, Eq. (\ref{conscharge}), consists of a fermionic as well as a bosonic component, thereby directly supporting the partitioning of the spectral weight in Fig. (\ref{spec}b).

\begin{figure}
\includegraphics[width = 8.5cm]{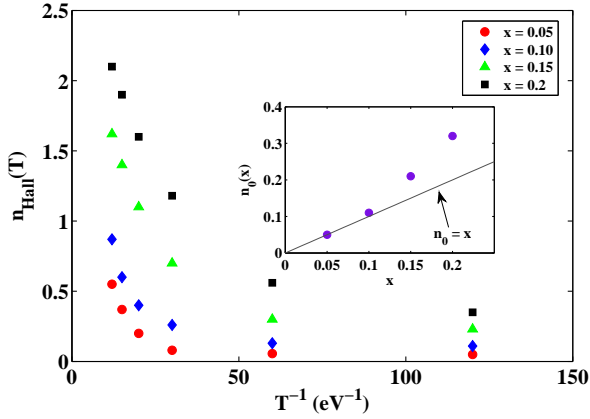}
\caption{n$_{\rm Hall}$ plotted as a function of inverse temperature for
  four different values of hole doping $x$: 1) solid circles,
  $x=0.05$, 2) diamonds, $x=0.10$, 3) triangles, $x=0.15$, and 4)
  squares, $x=0.2$. The inset shows the
  temperature independent part of the carrier density as a function of
  $x$.  Note it exceeds the nominal doping level indicated by the
  straight line. }
\label{fig1}
\end{figure}
We obtained the Green function $G(\bf{k},\omega)$ by numerical
integration and the subsequent spectral function, $A(\vec k,\omega)$.
We computed the Hall
coefficient from the spectral function using 
\beq
R_H = \sigma_{\rm xy}/\sigma_{\rm xx}^2,
\eeq
where
\beq
\sigma_{\rm xy}& =& \frac{2\pi^{2}|e|^{3}aB}{3\hbar^{2}}\int d\omega
(\frac{\partial f(\omega)}{\partial \omega})\frac{1}{N} \displaystyle
\sum_{\bf{k}}(\frac{\partial \epsilon_{\bf{k}}}{\partial{k_x}})^2
\nonumber\\
&\times&\frac{\partial^2 \epsilon_{\bf{k}}}{\partial {k_y}^2} A(\bf{k},\omega)^3 
\eeq
and
\beq
\sigma_{\rm xx} = \frac{\pi e^2 }{2 \hbar a} \int d\omega (-\frac{\partial f(\omega)}{\partial \omega})\frac{1}{N} \displaystyle \sum_{\bf{k}}(\frac{\partial \epsilon_{\bf{k}}}{\partial{k_x}})^2 A(\bf{k},\omega)^2
\eeq
with $\sigma_{\rm xx}$ and $\sigma_{\rm xy}$ the diagonal and off-diagonal
components of the conductivity tensor respectively, $f(\omega)$ is the
Fermi distribution function, and $B$ is the normal component of the
external magnetic field. The effective charge carrier density $n_{\rm
  Hall}$ is then obtained using the relation $R_H = -1/(n_{\rm Hall}
e)$. Fig.\ref{fig1} shows a set of plots of $n_{\rm Hall}$ as a
function of the inverse temperature, each corresponding to a different
value of hole-doping, $x$, in the underdoped regime ($x$ ranging from
0.05 to 0.20). The plots fit remarkably well to an exponentially
decaying form.  In other words, the computed charge carrier density
within the charge 2e boson theory of a doped Mott insulator
agrees well with the form given in Eq. (\ref{n}) proposed by Gor'kov
and Teitel'baum\cite{gorkov}. The inset shows the
temperature-independent part of the charge density as a function of
$x$.  This quantity exceeds the nominal doping level\cite{shenweakcoupling}.  Such a deviation the nominal doping level is also supported by angle-resolved photoemission spectroscopy (ARPES). Shown in Fig. (\ref{shenfs}) is a plot of the measured Fermi surface, $x_{\rm FS}$ in LSCO as a function of the nominal doping level $x$.  As is clear, $x_{\rm FS}$ deviates from linearity precisely where our calculated value of $n_0(x)$ does. In fact, the agreement between the inset in Fig. (\ref{fig1}) and 
Fig. (\ref{shenfs}) is striking.  Nonetheless, in previous 
publications\cite{shenweakcoupling}, this deviation was attributed to a band dispersion effect. The clear corroboration of this effect with our strong-coupling calculation suggests that the ultimate source of the deviation from linearity is Mottness. In a hole-doped Mott insulator, the Hall coefficient must change sign\cite{hall} before the particle-hole symmetric condition for the atomic limit, namely when $2x=1-x$ or equivalently before $x=1/3$\cite{stanescuhall,shastryhall}.  

\begin{figure}
\includegraphics[width = 8.5cm]{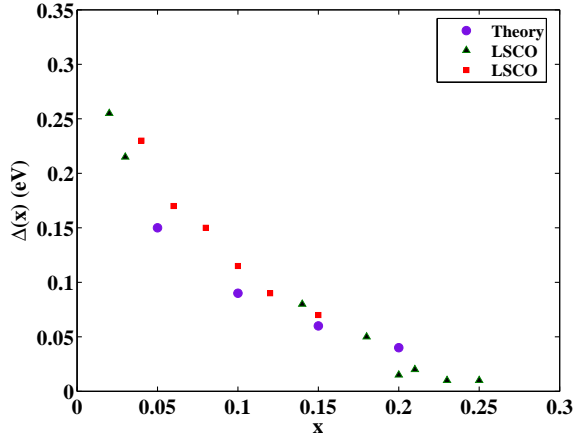}
\caption{$\Delta(x)$ (solid circles) obtained from fitting the plots in
  Fig.(\ref{fig1}) to Eq.(~\ref{n}) plotted as a function of hole
  doping $x$. The experimental values are also shown for LSCO: solid 
triangles\cite{andohall,onohall,delta1} and squares\cite{nishikawa}  The
  excellent agreement indicates that the bound component contributing
  to the charge density does in fact give rise to the pseudogap.}
\label{fig2}
\end{figure}

 \begin{figure}
\includegraphics[width = 8.5cm]{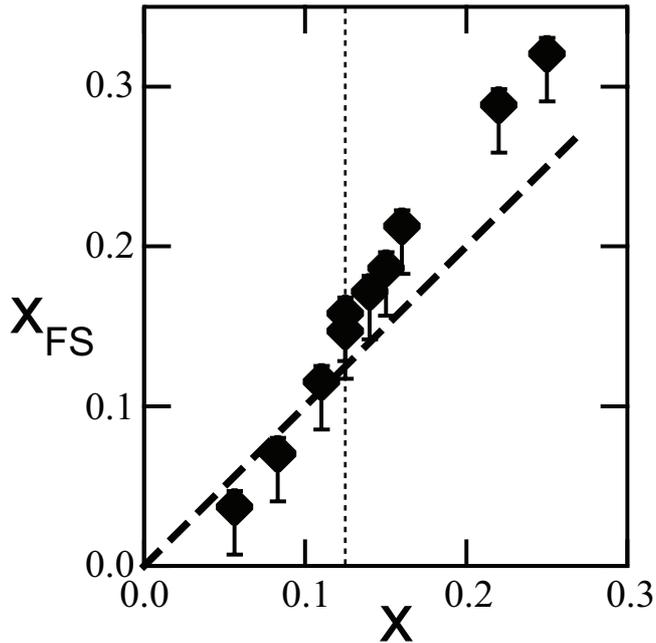}
\caption{Measured Fermi surface from ARPES (extracted from http://arxiv: 0911.2245) plotted versus the nominal doping level.  The deviation from linearity is corroborated by Fig. (\ref{fig2}) thereby signifying that strong coupling physics is the root cause.  }
\label{shenfs}
\end{figure}

\begin{figure}
\includegraphics[width = 9.0cm]{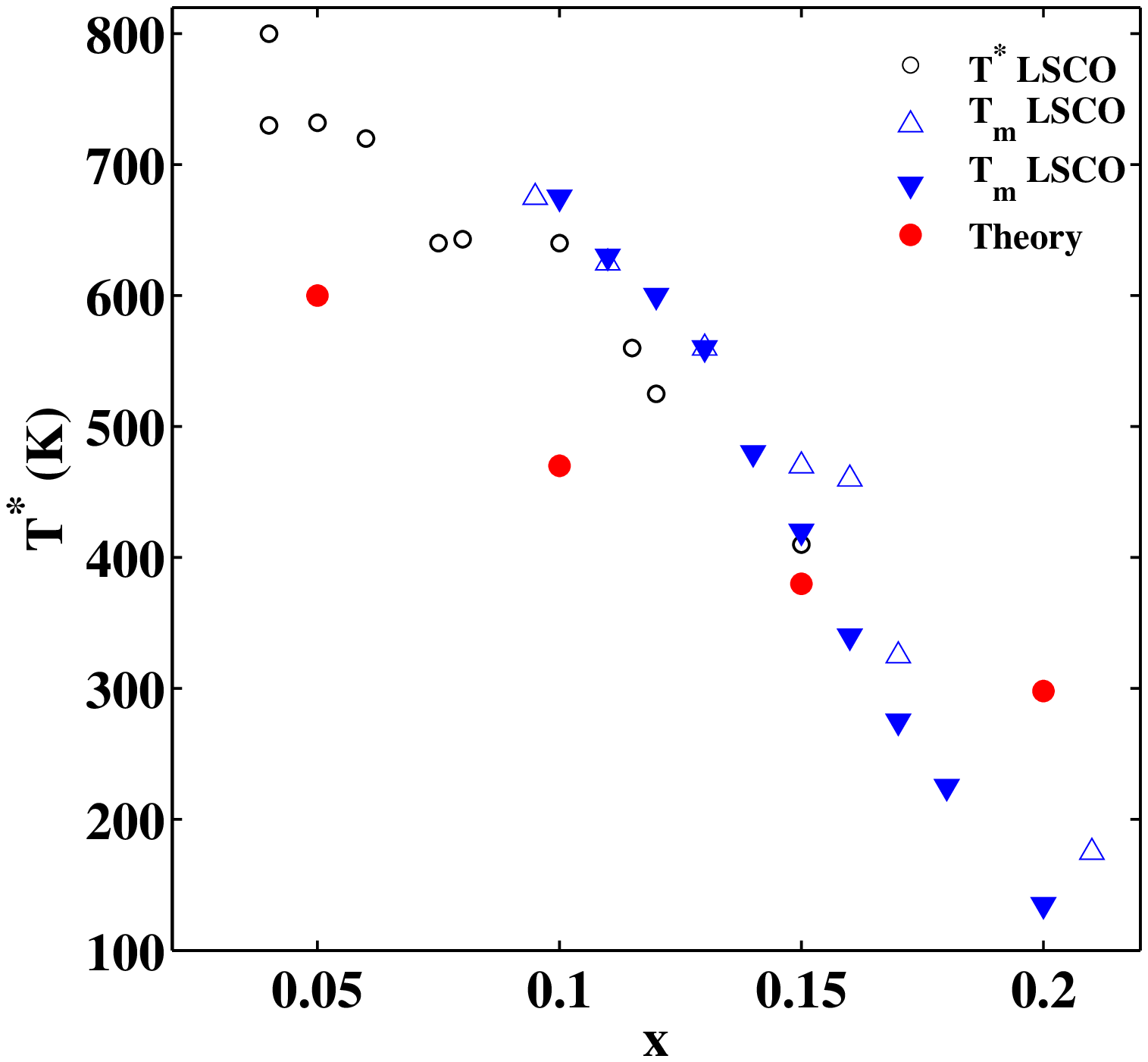}
\caption{$T^*(x))$ (solid circles) obtained from  Eq.(~\ref{T*})
  plotted as a function of hole doping $x$. The experimental data were
  gleaned from the following: open circles are from Ref.\cite{timusk}$T^*$,  open triangles ($T_m$) from Ref.\cite{yoshizaki}, and closed triangles ($T_m$) from Ref.\cite{nakano}. }
\label{fig3}
\end{figure}

The `binding energy', $\Delta(x)$, was extracted for each doping and plotted in
 Fig.(\ref{fig2}) using Eq. (\ref{n}).  Shown here also are the values
 for the experimentally determined pseudogap energy for LSCO\cite{andohall,onohall,delta1}.  The
 magnitude of $\Delta(x)$ falls with increasing hole doping as is seen
 experimentally and hence is consistent with its interpretation, even quantitatively, as a
 measure of the pseudogap temperature $T^*$. A rough estimate
 of $T^*$,
 \beq
 \label{T*}
 T^*(x)\approx -\Delta(x)/\ln(x),
 \eeq
may be obtained from $\Delta(x)$, by equating the number of doped
carriers $x$ with that of the activated ones
$n_{1}(x)exp(-\Delta(x,T))$ as proposed by Gor'kov and Teitel'baum\cite{gorkov}.
Fig.(\ref{fig3}) shows a plot of $T^*$ as a function of $x$. This
result is in qualitative agreement with the  experimentally obtained estimates of
$T^*$\cite{timusk,yoshizaki,nakano}.  

\subsection{T-linear Resistivity}

The physical picture for the charge 2e boson calculation is now
clear.  Below a characteristic temperature the boson is bound to a
hole and produces charge $e$ states.  This leads to a non-zero value
for $\alpha$.  Above, $T^\ast$, the bound states break up. The
simplest way of understanding why the charge 2e boson must be bound at
low energies, aside from the fact that it has no bare dynamics, is that once the high-energy sector is integrated out
exactly, the Hilbert space shrinks back to the Fock space of the
Hubbard model.  

\begin{figure}
\centering
\includegraphics[width=7.5cm]{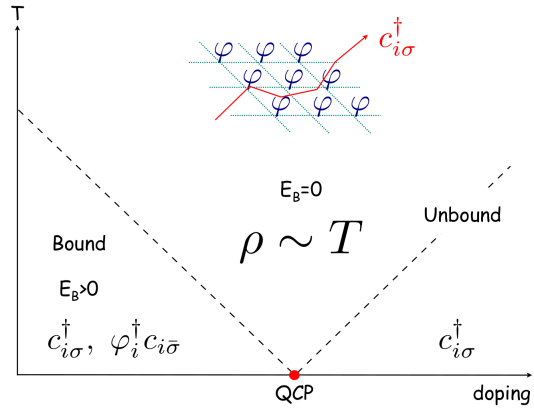}
\caption{Phase diagram for the dynamics mediated by the charge 2e boson, $\varphi_i$.  Bound states between form between the holes and the charge $2e$ boson as seen in the calculation of the Hall coefficient, giving rise to $E_B>0$ with $E_B$ the binding energy.  The pseudogap regime terminates at a quantum critical point (QCP) where the bosons and holes  unbind. In the critical regime, the dominant scattering mechanism is still due to the interaction with the charge $2e$ bosons. In this regime, the energy to excite the bosons vanishes.  T-linear resistivity results anytime electrons scatter off classical bosons with $T>\omega_b$, where $\omega_b$ is the energy to excite a boson as in the electron-phonon problem above the Debye temperature. To the right of the quantum critical regime, the boson is irrelevant and scattering is dominated by electron-electron interactions indicative of a Fermi liquid.}\label{tlin}
\end{figure}

The mechanism for $T-$linear resistivity is simple within
this model. Once the binding energy of the boson vanishes, bosons are
free to scatter off the electrons. The absence of a kinetic energy
term for the bosons implies that their dynamics are classical.  The resistivity of electrons
scattering off classical bosons is well-known to scale
 linearly with temperature above the
energy to create the boson as depicted in Fig. (\ref{tlin}).  Hence, this mechanism is robust and should
persist to high temperatures.   Consequently, the charge 2e boson
reduction of the Hubbard model offers a resolution of the pseudogap and the transition to the
strange metal regime of the cuprates.   Further, the
mechanism for the strange metal is
consistent with the scaling analysis leading to Eq. (\ref{dclimit}).  Namely,
$T-$linear resistivity requires an additional energy scale absent from
a single-parameter scaling analysis.  In the exact low-energy theory, a charge 2e boson
emerges as a new degree of freedom.  While it is bound in the
pseudogap regime, its unbinding beyond a critical temperature or
doping provides the added degree of freedom to generate the anomalous
temperature dependence for the resistivity.  A further experimental
prediction of this work then is that the strange metal regime should
be populated with charge 2e excitations, without the usual diamagnetic
signal.  Shot noise measurements are ideally suited for testing this
prediction.  

\section{Final Remarks}

We have focused here on explaining three inter-related facts.  F1) Breaking the Landau correspondence between the Fermi gas and the
interacting system requires new degrees of freedom.  F2)   The connection
between $T-$linear resistivity and quantum criticality is only
possible if an additional degree of freedom outside the one-parameter
paradigm is present as Eq. (\ref{dclimit}) lays plain.  F3) The total weight of the low-energy band
exceeds the number of ways electrons can be assigned to this band,
thereby requiring new low-energy degrees of freedom.  The common
ingredient among F1-F3 is a new degree of freedom distinct from the electrons. Explicitly
carrying out the Wilsonian program for the Hubbard model by
integrating out the high-energy sector produces the missing degree of freedom that is
capable of explaining all of these facts.    Since it is the mixing with the doubly
occupied sector that enhances the intensity of the LHB beyond $1+x$,
the new degree of freedom at low energies must have charge $2e$ and
hence must be bosonic. This has been explicitly demonstrated by the
exact procedure to integrate out the high-energy sector.  At
low energies and temperatures, the charge $2e$ boson mediates new
charge $e$ states in the LHB.  Such bound states already exist in the
half-filled system and constitute the propagating degrees of freedom
that describe the UHB and LHB.  Hence, the pseudogap emerges in the
charge 2e theory as a remnant of the Mott gap.  Heuristically, the bound
states and hence the gap originate from a doubly
occupied site bound to a hole.

This work makes a series of experimental predictions.  First, below the
$T^\ast$ line $L/n_h>1$ whereas above $L/n_h=1$.
This information can be extracted from temperature-dependent experiments of the kind leading to
Fig. (\ref{mottcollapse}) .  Second, any experimental probe that couples
to the low-energy excitations should be interpreted in terms of $x'$,
not the bare hole number $x$.  These include measurements of the 1)
optical conductivity, 2) 
superfluid density, and 3) `fermi surface' volumes extracted from
quantum oscillation experiments (typically Hall measurements). The latter is particularly germane because
the Fermi surface
volumes extracted experimentally\cite{qoscill,osc1} for YBCO are not consistent with any integer
multiple of the physically doped holes.

That the charge $2e$ boson mediates
local bound states is {\it a priori} expected as it lacks a bare
kinetic term.  The bound states that form account for the pseudogap
and their breakup leads to a resistivity that quite generally scales
linearly with temperature.  As long as the UHB is present, the program
carried out here is sufficient to describe the physics of a doped Mott
insulator. Once the UHB collapses, traditional Fermi liquid
descriptions obtain.  
As a separation between the UHB and LHB is only meaningful if there is
a gap in the spectrum, the pseudogap lies at the heart of
Mottness. Precisely the role such bound states play in the
superconducting state remains open.

\acknowledgements{I thank R. Leigh, T.-P. Choy, S. Hong, S. Chakraborty, D. Galanakis, and S. Chakraborty for extensive collaborations which led to the work summarized here. I also thank Frank Kr\"uger, Lance Cooper and P. Abbamontte for extensive discussions on Mottness. I also acknowledge financial support from the NSF DMR-0940992 and the Center for Emergent Superconductivity, a DoE Energy Frontier Research Center, Award Number DE-AC0298CH1088 .}

\newcommand{\noopsort}[1]{} \newcommand{\printfirst}[2]{#1}
  \newcommand{\singleletter}[1]{#1} \newcommand{\switchargs}[2]{#2#1}


\begin{thebibliography}{93}
\providecommand{\natexlab}[1]{#1}
\expandafter\ifx\csname urlstyle\endcsname\relax
  \providecommand{\doi}[1]{doi:\discretionary{}{}{}#1}\else
  \providecommand{\doi}{doi:\discretionary{}{}{}\begingroup
  \urlstyle{rm}\Url}\fi

\bibitem[{Abbamonte \emph{et~al.}(2005)Abbamonte, Rusydi, Smadici, Gu, Sawatzky
  \& Feng}]{stripes4}
Abbamonte, P., Rusydi, A., Smadici, S., Gu, G.~D., Sawatzky, G.~A. \& Feng,
  D.~L. 2005 Spatially modulated 'mottness' in la2-xbaxcuo4.
\newblock \emph{Nat Phys}, \textbf{1}(3), 155--158.

\bibitem[{Alloul \emph{et~al.}(1989)Alloul, Ohno \& Mendels}]{pg}
Alloul, H., Ohno, T. \& Mendels, P. 1989 $y89$ nmr evidence for a fermi-liquid
  behavior in $yba2$$cu3$$o6+x$.
\newblock \emph{Phys. Rev. Lett.}, \textbf{63}(16), 1700--1703.
\newblock (\doi{10.1103/PhysRevLett.63.1700})

\bibitem[{Anderson(1987)}]{rvb}
Anderson, P.~W. 1987 The resonating valence bond state in la2cuo4 and
  superconductivity.
\newblock \emph{Science}, \textbf{235}(4793), 1196--1198.

\bibitem[{Ando \emph{et~al.}(2004{\natexlab{\emph{a}}})Ando, Komiya, Segawa,
  Ono \& Kurita}]{ando}
Ando, Y., Komiya, S., Segawa, K., Ono, S. \& Kurita, Y.
  2004{\natexlab{\emph{a}}} Electronic phase diagram of high-$tc$ cuprate
  superconductors from a mapping of the in-plane resistivity curvature.
\newblock \emph{Phys. Rev. Lett.}, \textbf{93}(26), 267\,001.
\newblock (\doi{10.1103/PhysRevLett.93.267001})

\bibitem[{Ando \emph{et~al.}(2004{\natexlab{\emph{b}}})Ando, Kurita, Komiya,
  Ono \& Segawa}]{andohall}
Ando, Y., Kurita, Y., Komiya, S., Ono, S. \& Segawa, K.
  2004{\natexlab{\emph{b}}} Evolution of the hall coefficient and the peculiar
  electronic structure of the cuprate superconductors.
\newblock \emph{Physical Review Letters}, \textbf{92}(19).

\bibitem[{Balents \emph{et~al.}(1999)Balents, Fisher \& Nayak}]{nodalmetal}
Balents, L., Fisher, M. P.~A. \& Nayak, C. 1999 Dual order parameter for the
  nodal liquid.
\newblock \emph{Physical Review B}, \textbf{60}(3).

\bibitem[{Benfatto \& Gallavotti(1990)}]{others}
Benfatto, G. \& Gallavotti, G. 1990 Perturbation theory of the fermi surface in
  a quantum liquid: A general quasiparticle formalism and one-dimensional
  systems.
\newblock \emph{Journal of Statistical Physics}, \textbf{959}, 541--664.

\bibitem[{Castellani \emph{et~al.}(1979)Castellani, Castro, Feinberg \&
  Ranninger}]{castellani}
Castellani, C., Castro, C.~D., Feinberg, D. \& Ranninger, J. 1979 New model
  hamiltonian for the metal-insulator transition.
\newblock \emph{Phys. Rev. Lett.}, \textbf{43}(26), 1957--1960.
\newblock (\doi{10.1103/PhysRevLett.43.1957})

\bibitem[{Chakraborty \emph{et~al.}(2009)Chakraborty, Hong \&
  Phillips}]{dswtfinal1}
Chakraborty, S., Hong, S. \& Phillips, P. 2009 Non-conservation of fermionic
  degrees of freedom in doped mott insulators.
\newblock \emph{arXiv:0909.2854}.

\bibitem[{Chakraborty \& Phillips(2009)}]{hallcalc}
Chakraborty, S. \& Phillips, P. 2009 Two-fluid model of the pseudogap of
  high-temperature cuprate superconductors based on charge-2e bosons.
\newblock \emph{Physical Review B (Condensed Matter and Materials Physics)},
  \textbf{80}(13), 132505.
\newblock (\doi{10.1103/PhysRevB.80.132505})

\bibitem[{Chakravarty \emph{et~al.}(2001)Chakravarty, Laughlin, Morr \&
  Nayak}]{ddw}
Chakravarty, S., Laughlin, R.~B., Morr, D.~K. \& Nayak, C. 2001 Hidden order in
  the cuprates.
\newblock \emph{Phys. Rev. B}, \textbf{63}(9), 094\,503.
\newblock (\doi{10.1103/PhysRevB.63.094503})

\bibitem[{Chen \emph{et~al.}(1992)Chen, Tjeng, Kwo, Kao, Rudolf, Sette \&
  Fleming}]{chen}
Chen, C.~T., Tjeng, L.~H., Kwo, J., Kao, H.~L., Rudolf, P., Sette, F. \&
  Fleming, R.~M. 1992 Out-of-plane orbital characters of intrinsic and doped
  holes in $la2-x$$srx$$cuo4$.
\newblock \emph{Phys. Rev. Lett.}, \textbf{68}(16), 2543--2546.
\newblock (\doi{10.1103/PhysRevLett.68.2543})

\bibitem[{Choy \emph{et~al.}(2008{\natexlab{\emph{a}}})Choy, Leigh \&
  Phillips}]{charge2e2}
Choy, T.-P., Leigh, R.~G. \& Phillips, P. 2008{\natexlab{\emph{a}}} Hidden
  charge-2e boson: Experimental consequences for doped mott insulators.
\newblock \emph{Physical Review B (Condensed Matter and Materials Physics)},
  \textbf{77}(10), 104\,524--9.

\bibitem[{Choy \emph{et~al.}(2008{\natexlab{\emph{b}}})Choy, Leigh, Phillips \&
  Powell}]{charge2e1}
Choy, T.-P., Leigh, R.~G., Phillips, P. \& Powell, P.~D.
  2008{\natexlab{\emph{b}}} Exact integration of the high energy scale in doped
  mott insulators.
\newblock \emph{Physical Review B (Condensed Matter and Materials Physics)},
  \textbf{77}(1), 014\,512--12.

\bibitem[{Daou \emph{et~al.}(2009)Daou, Chang, LeBoeuf, Cyr-Choiniere,
  Laliberte, Doiron-Leyraud, Ramshaw, Liang, Bonn \emph{et~al.}}]{bonnnematic}
Daou, R., Chang, J., LeBoeuf, D., Cyr-Choiniere, O., Laliberte, F.,
  Doiron-Leyraud, N., Ramshaw, B.~J., Liang, R., Bonn, D.~A. \emph{et~al.} 2009
  Broken rotational symmetry in the pseudogap phase of a high-tc
  superconductor.
\newblock \emph{arXiv:0909.4430}.

\bibitem[{Doiron-Leyraud \emph{et~al.}(2007)Doiron-Leyraud, Proust, LeBoeuf,
  Levallois, Bonnemaison, Liang, Bonn, Hardy \& Taillefer}]{qoscill}
Doiron-Leyraud, N., Proust, C., LeBoeuf, D., Levallois, J., Bonnemaison, J.-B.,
  Liang, R., Bonn, D.~A., Hardy, W.~N. \& Taillefer, L. 2007 Quantum
  oscillations and the fermi surface in an underdoped high-tc superconductor.
\newblock \emph{Nature}, \textbf{447}(7144), 565--568.

\bibitem[{Drechsler \emph{et~al.}(1997)Drechsler, M{\'a}lek \&
  Eschrig}]{drechsler}
Drechsler, S.~L., M{\'a}lek, J. \& Eschrig, H. 1997 Exact diagonalization study
  of the hole distribution in cuo3 chains within the four-band dp model.
\newblock \emph{Physical Review B}, \textbf{55}(1).

\bibitem[{Dzyaloshinskii(2003)}]{zeros1}
Dzyaloshinskii, I. 2003 Some consequences of the luttinger theorem: The
  luttinger surfaces in non-fermi liquids and mott insulators.
\newblock \emph{Phys. Rev. B}, \textbf{68}(8), 085\,113.
\newblock (\doi{10.1103/PhysRevB.68.085113})

\bibitem[{Eskes \emph{et~al.}(1994)Eskes, Ole\ifmmode~\acute{s}\else
  \'{s}\fi{}, Meinders \& Stephan}]{eskes}
Eskes, H., Ole\ifmmode~\acute{s}\else \'{s}\fi{}, A.~M., Meinders, M. B.~J. \&
  Stephan, W. 1994 Spectral properties of the hubbard bands.
\newblock \emph{Phys. Rev. B}, \textbf{50}(24), 17\,980--18\,002.
\newblock (\doi{10.1103/PhysRevB.50.17980})

\bibitem[{Essler \& Tsvelik(2002)}]{zeros2}
Essler, F. H.~L. \& Tsvelik, A.~M. 2002 Weakly coupled one-dimensional mott
  insulators.
\newblock \emph{Phys. Rev. B}, \textbf{65}(11), 115\,117.
\newblock (\doi{10.1103/PhysRevB.65.115117})

\bibitem[{Faulkner \emph{et~al.}(2009)Faulkner, Liu, McGreevy \& Vegh}]{lmv}
Faulkner, T., Liu, H., McGreevy, J. \& Vegh, D. 2009 Emergent quantum
  criticality, fermi surfaces, and ads2.
\newblock \emph{arXiv:0907.2694}.

\bibitem[{Fauque \emph{et~al.}(2006)Fauque, Sidis, Hinkov, Pailhes, Lin, Chaud
  \& Bourges}]{trsb4}
Fauque, B., Sidis, Y., Hinkov, V., Pailhes, S., Lin, C.~T., Chaud, X. \&
  Bourges, P. 2006 Magnetic order in the pseudogap phase of high-t{$[$}sub
  c{$]$} superconductors.
\newblock \emph{Physical Review Letters}, \textbf{96}(19), 197\,001--4.

\bibitem[{Franz \& Te\ifmmode \check{s}\else
  \v{s}\fi{}anovi\ifmmode~\acute{c}\else \'{c}\fi{}(2001)}]{inco3}
Franz, M. \& Te\ifmmode \check{s}\else \v{s}\fi{}anovi\ifmmode~\acute{c}\else
  \'{c}\fi{}, Z. 2001 Algebraic fermi liquid from phase fluctuations:
  \char16{}topological\char17{} fermions, vortex \char16{}berryons,\char17{}
  and $qed3$ theory of cuprate superconductors.
\newblock \emph{Phys. Rev. Lett.}, \textbf{87}(25), 257\,003.
\newblock (\doi{10.1103/PhysRevLett.87.257003})

\bibitem[{Georges \emph{et~al.}(1996)Georges, Kotliar, Krauth \&
  Rozenberg}]{dinfty}
Georges, A., Kotliar, G., Krauth, W. \& Rozenberg, M.~J. 1996 Dynamical
  mean-field theory of strongly correlated fermion systems and the limit of
  infinite dimensions.
\newblock \emph{Rev. Mod. Phys.}, \textbf{68}(1), 13.
\newblock (\doi{10.1103/RevModPhys.68.13})

\bibitem[{Gor'kov \& Teitel'baum(2006)}]{gorkov}
Gor'kov, L.~P. \& Teitel'baum, G.~B. 2006 Interplay of externally doped and
  thermally activated holes in la{$[$}sub 2-x{$]$}sr{$[$}sub x{$]$}cuo{$[$}sub
  4{$]$} and their impact on the pseudogap crossover.
\newblock \emph{Physical Review Letters}, \textbf{97}(24), 247\,003--4.

\bibitem[{Harris \& Lange(1967)}]{hl}
Harris, A.~B. \& Lange, R.~V. 1967 Single-particle excitations in narrow energy
  bands.
\newblock \emph{Phys. Rev.}, \textbf{157}(2), 295--314.
\newblock (\doi{10.1103/PhysRev.157.295})

\bibitem[{Haug \emph{et~al.}(2009)Haug, Hinkov, Suchaneck, Inosov, Christensen,
  Niedermayer, Bourges, Sidis, Park \emph{et~al.}}]{keimernematic}
Haug, D., Hinkov, V., Suchaneck, A., Inosov, D.~S., Christensen, N.~B.,
  Niedermayer, C., Bourges, P., Sidis, Y., Park, J.~T. \emph{et~al.} 2009
  Magnetic-field-enhanced incommensurate magnetic order in the underdoped
  high-temperature superconductor yba[sub 2]cu[sub 3][bold o][sub 6.45].
\newblock \emph{Physical Review Letters}, \textbf{103}(1), 017001.
\newblock (\doi{10.1103/PhysRevLett.103.017001})

\bibitem[{Honma \& Hor(2008)}]{honma}
Honma, T. \& Hor, P.~H. 2008 Unified electronic phase diagram for hole-doped
  high-t{$[$}sub c{$]$} cuprates.
\newblock \emph{Physical Review B (Condensed Matter and Materials Physics)},
  \textbf{77}(18), 184\,520--16.

\bibitem[{Howson \emph{et~al.}(1990)Howson, Salamon, Friedmann, Rice \&
  Ginsberg}]{salamon}
Howson, M.~A., Salamon, M.~B., Friedmann, T.~A., Rice, J.~P. \& Ginsberg, D.
  1990 Anomalous peak in the thermopower of
  yba{\_}{\{}2{\}}cu{\_}{\{}3{\}}o{\_}{\{}7- delta {\}} single crystals: A
  possible fluctuation effect.
\newblock \emph{Physical Review B}, \textbf{41}(1).

\bibitem[{Hufner \emph{et~al.}(2008)Hufner, Hossain, Damascelli \&
  Sawatzky}]{pg2gaps}
Hufner, S., Hossain, M.~A., Damascelli, A. \& Sawatzky, G.~A. 2008 Two gaps
  make a high-temperature superconductor?
\newblock \emph{Reports on Progress in Physics}, \textbf{71}(6), 062\,501
  (9pp).

\bibitem[{Hussey \emph{et~al.}(2003)Hussey, Abdel-Jawad, Carrington, Mackenzie
  \& Balicas}]{amro}
Hussey, N.~E., Abdel-Jawad, M., Carrington, A., Mackenzie, A.~P. \& Balicas, L.
  2003 A coherent three-dimensional fermi surface in a
  high-transition-temperature superconductor.
\newblock \emph{Nature}, \textbf{425}(6960), 814--817.

\bibitem[{Hybertsen \emph{et~al.}(1992)Hybertsen, Stechel, Foulkes \&
  Schl{\"u}ter}]{stechel}
Hybertsen, M.~S., Stechel, E.~B., Foulkes, W. M.~C. \& Schl{\"u}ter, M. 1992
  Model for low-energy electronic states probed by x-ray absorption in high-tc
  cuprates.
\newblock \emph{Physical Review B}, \textbf{45}(17).

\bibitem[{Ino \emph{et~al.}(2002)Ino, Kim, Nakamura, Yoshida, Mizokawa,
  Fujimori, Shen, Kakeshita, Eisaki \emph{et~al.}}]{srd214a}
Ino, A., Kim, C., Nakamura, M., Yoshida, T., Mizokawa, T., Fujimori, A., Shen,
  Z.~X., Kakeshita, T., Eisaki, H. \emph{et~al.} 2002 Doping-dependent
  evolution of the electronic structure of la2-xsrxcuo4 in the superconducting
  and metallic phases.
\newblock \emph{Physical Review B}, \textbf{65}(9).

\bibitem[{Kaminski \emph{et~al.}(2002)Kaminski, Rosenkranz, Fretwell,
  Campuzano, Li, Raffy, Cullen, You, Olson \emph{et~al.}}]{trsb3}
Kaminski, A., Rosenkranz, S., Fretwell, H.~M., Campuzano, J.~C., Li, Z., Raffy,
  H., Cullen, W.~G., You, H., Olson, C.~G. \emph{et~al.} 2002 Spontaneous
  breaking of time-reversal symmetry in the pseudogap state of a high-tc
  superconductor.
\newblock \emph{Nature}, \textbf{416}(6881), 610--613.

\bibitem[{Kanigel \emph{et~al.}(2006)Kanigel, Norman, Randeria, Chatterjee,
  Souma, Kaminski, Fretwell, Rosenkranz, Shi \emph{et~al.}}]{kanigel}
Kanigel, A., Norman, M.~R., Randeria, M., Chatterjee, U., Souma, S., Kaminski,
  A., Fretwell, H.~M., Rosenkranz, S., Shi, M. \emph{et~al.} 2006 Evolution of
  the pseudogap from fermi arcs to the nodal liquid.
\newblock \emph{Nat Phys}, \textbf{2}(7), 447--451.

\bibitem[{Kivelson \emph{et~al.}(1998)Kivelson, Fradkin \& Emery}]{stripes1}
Kivelson, S.~A., Fradkin, E. \& Emery, V.~J. 1998 Electronic liquid-crystal
  phases of a doped mott insulator.
\newblock \emph{Nature}, \textbf{393}(6685), 550--553.

\bibitem[{Kohn(1964)}]{kohn1}
Kohn, W. 1964 Theory of the insulating state.
\newblock \emph{Physical Review}, \textbf{133}(1A).

\bibitem[{Kondo \emph{et~al.}(2005)Kondo, Takeuchi, Mizutani, Yokoya, Tsuda \&
  Shin}]{arp1}
Kondo, T., Takeuchi, T., Mizutani, U., Yokoya, T., Tsuda, S. \& Shin, S. 2005
  Contribution of electronic structure to thermoelectric power in $ (bi,pb)2
  (sr,la)2 cu o6+\delta{} $.
\newblock \emph{Phys. Rev. B}, \textbf{72}(2), 024\,533.
\newblock (\doi{10.1103/PhysRevB.72.024533})

\bibitem[{Kotegawa \emph{et~al.}(2001)Kotegawa, Tokunaga, Ishida, Zheng,
  Kitaoka, Kito, Iyo, Tokiwa, Watanabe \emph{et~al.}}]{NQR}
Kotegawa, H., Tokunaga, Y., Ishida, K., Zheng, G.-q., Kitaoka, Y., Kito, H.,
  Iyo, A., Tokiwa, K., Watanabe, T. \emph{et~al.} 2001 Unusual magnetic and
  superconducting characteristics in multilayered high-$tc$ cuprates: $63cu$
  nmr study.
\newblock \emph{Phys. Rev. B}, \textbf{64}(6), 064\,515.
\newblock (\doi{10.1103/PhysRevB.64.064515})

\bibitem[{Kyung \emph{et~al.}(2006)Kyung, Kancharla, S\'en\'echal, Tremblay,
  Civelli \& Kotliar}]{kyung}
Kyung, B., Kancharla, S.~S., S\'en\'echal, D., Tremblay, A.-M.~S., Civelli, M.
  \& Kotliar, G. 2006 Pseudogap induced by short-range spin correlations in a
  doped mott insulator.
\newblock \emph{Phys. Rev. B}, \textbf{73}(16), 165\,114.
\newblock (\doi{10.1103/PhysRevB.73.165114})

\bibitem[{Le~Tacon \emph{et~al.}(2006)Le~Tacon, Sacuto, Georges, Kotliar,
  Gallais, Colson \& Forget}]{0603392}
Le~Tacon, M., Sacuto, A., Georges, A., Kotliar, G., Gallais, Y., Colson, D. \&
  Forget, A. 2006 Two energy scales and two distinct quasiparticle dynamics in
  the superconducting state of underdoped cuprates.
\newblock \emph{Nat Phys}, \textbf{2}(8), 537--543.

\bibitem[{LeBoeuf \emph{et~al.}(2007)LeBoeuf, Doiron-Leyraud, Levallois, Daou,
  Bonnemaison, Hussey, Balicas, Ramshaw, Liang \emph{et~al.}}]{osc1}
LeBoeuf, D., Doiron-Leyraud, N., Levallois, J., Daou, R., Bonnemaison, J.~B.,
  Hussey, N.~E., Balicas, L., Ramshaw, B.~J., Liang, R. \emph{et~al.} 2007
  Electron pockets in the fermi surface of hole-doped high-tc superconductors.
\newblock \emph{Nature}, \textbf{450}(7169), 533--536.

\bibitem[{Leigh \& Phillips(2008)}]{charge2e4}
Leigh, R.~G. \& Phillips, P. 2008 Origin of the mott gap.
\newblock \emph{arXiv:0812.0593}.

\bibitem[{Leigh \emph{et~al.}(2007)Leigh, Phillips \& Choy}]{charge2e}
Leigh, R.~G., Phillips, P. \& Choy, T.-P. 2007 Hidden charge 2e boson in doped
  mott insulators.
\newblock \emph{Physical Review Letters}, \textbf{99}(4), 046\,404--4.

\bibitem[{Liang \emph{et~al.}(2006)Liang, Bonn \& Hardy}]{hardy}
Liang, R., Bonn, D.~A. \& Hardy, W.~N. 2006 Evaluation of cuo{$[$}sub 2{$]$}
  plane hole doping in yba{$[$}sub 2{$]$}cu{$[$}sub 3{$]$}o{$[$}sub 6 + x{$]$}
  single crystals.
\newblock \emph{Physical Review B (Condensed Matter and Materials Physics)},
  \textbf{73}(18), 180\,505--4.

\bibitem[{Liebsch(2010)}]{liebsch}
Liebsch, A. 2010 Spectral weight of doping-induced states in the 2d hubbard
  model.
\newblock \emph{arXiv:1004.1322}.

\bibitem[{Machida(1989)}]{stripes6}
Machida, K. 1989 Magnetism in la2cuo4 based compounds.
\newblock \emph{Physica C: Superconductivity}, \textbf{158}(1-2), 192--196.

\bibitem[{Maldacena(1998)}]{maldacena}
Maldacena, J. 1998 The large n limit of superconformal field theories and
  supergravity.
\newblock \emph{Adv. Theor. Math. Phys.}, \textbf{2}, 231--252.

\bibitem[{Markiewicz \& Kusko(2002)}]{markiewicz}
Markiewicz, R.~S. \& Kusko, C. 2002 Phase separation models for cuprate stripe
  arrays.
\newblock \emph{Physical Review B}, \textbf{65}(6).

\bibitem[{Meinders \emph{et~al.}(1993)Meinders, Eskes \& Sawatzky}]{sawatzky}
Meinders, M. B.~J., Eskes, H. \& Sawatzky, G.~A. 1993 Spectral-weight transfer:
  Breakdown of low-energy-scale sum rules in correlated systems.
\newblock \emph{Phys. Rev. B}, \textbf{48}(6), 3916--3926.
\newblock (\doi{10.1103/PhysRevB.48.3916})

\bibitem[{Merz \emph{et~al.}(1998{\natexlab{\emph{a}}})Merz, N{\"u}cker,
  Schweiss, Schuppler, Chen, Chakarian, Freeland, Idzerda, Kl{\"a}ser
  \emph{et~al.}}]{merz}
Merz, M., N{\"u}cker, N., Schweiss, P., Schuppler, S., Chen, C.~T., Chakarian,
  V., Freeland, J., Idzerda, Y.~U., Kl{\"a}ser, M. \emph{et~al.}
  1998{\natexlab{\emph{a}}} Site-specific x-ray absorption spectroscopy of
  y1-xcaxba2cu3o7-y: Overdoping and role of apical oxygen for high temperature
  superconductivity.
\newblock \emph{Physical Review Letters}, \textbf{80}(23).

\bibitem[{Merz \emph{et~al.}(1998{\natexlab{\emph{b}}})Merz, N\"ucker,
  Schweiss, Schuppler, Chen, Chakarian, Freeland, Idzerda, Kl\"aser
  \emph{et~al.}}]{ody123}
Merz, M., N\"ucker, N., Schweiss, P., Schuppler, S., Chen, C.~T., Chakarian,
  V., Freeland, J., Idzerda, Y.~U., Kl\"aser, M. \emph{et~al.}
  1998{\natexlab{\emph{b}}} Site-specific x-ray absorption spectroscopy of
  $y1-xcaxba2cu3o7-y$: Overdoping and role of apical oxygen for high
  temperature superconductivity.
\newblock \emph{Phys. Rev. Lett.}, \textbf{80}(23), 5192--5195.
\newblock (\doi{10.1103/PhysRevLett.80.5192})

\bibitem[{Mott(1949)}]{mott}
Mott, N.~F. 1949 The basis of the electron theory of metals, with special
  reference to the transition metals.
\newblock \emph{Proceedings of the Physical Society. Section A},
  \textbf{62}(7), 416--422.

\bibitem[{Nakano \emph{et~al.}(1994)Nakano, Oda, Manabe, Momono, Miura \&
  Ido}]{nakano}
Nakano, T., Oda, M., Manabe, C., Momono, N., Miura, Y. \& Ido, M. 1994 Magnetic
  properties and electronic conduction of superconducting $la2-x$$srx$$cuo4$.
\newblock \emph{Phys. Rev. B}, \textbf{49}(22), 16\,000--16\,008.
\newblock (\doi{10.1103/PhysRevB.49.16000})

\bibitem[{Nishikawa \emph{et~al.}(1994)Nishikawa, Takeda \& Sato}]{nishikawa}
Nishikawa, T., Takeda, J. \& Sato, M. 1994 Transport anomalies of high-$t_{\rm
  c}$ oxides above room temperature.
\newblock \emph{Journal of the Physical Society of Japan}, \textbf{63}(4),
  1441--1448.

\bibitem[{Norman \emph{et~al.}(1998)Norman, Ding, Randeria, Campuzano, Yokoya,
  Takeuchi, Takahashi, Mochiku, Kadowaki \emph{et~al.}}]{norman}
Norman, M.~R., Ding, H., Randeria, M., Campuzano, J.~C., Yokoya, T., Takeuchi,
  T., Takahashi, T., Mochiku, T., Kadowaki, K. \emph{et~al.} 1998 Destruction
  of the fermi surface in underdoped high-tc superconductors.
\newblock \emph{Nature}, \textbf{392}(6672), 157--160.

\bibitem[{Norman \emph{et~al.}(2005)Norman, Pines \& Kallin}]{npk}
Norman, M.~R., Pines, D. \& Kallin, C. 2005 The pseudogap: Friend or foe of
  high tc?
\newblock \emph{Advances in Physics}, \textbf{54}, 715--733.

\bibitem[{N{\"u}cker \emph{et~al.}(1995)N{\"u}cker, Pellegrin, Schweiss, Fink,
  Molodtsov, Simmons, Kaindl, Frentrup, Erb \emph{et~al.}}]{nucker}
N{\"u}cker, N., Pellegrin, E., Schweiss, P., Fink, J., Molodtsov, S.~L.,
  Simmons, C.~T., Kaindl, G., Frentrup, W., Erb, A. \emph{et~al.} 1995
  Site-specific and doping-dependent electronic structure of
  yba{\_}{\{}2{\}}cu{\_}{\{}3{\}}o{\_}{\{}x{\}} probed by o 1s and cu 2p
  x-ray-absorption spectroscopy.
\newblock \emph{Physical Review B}, \textbf{51}(13).

\bibitem[{Ono \emph{et~al.}(2007)Ono, Komiya \& Ando}]{onohall}
Ono, S., Komiya, S. \& Ando, Y. 2007 Strong charge fluctuations manifested in
  the high-temperature hall coefficient of high-t{$[$}sub c{$]$} cuprates.
\newblock \emph{Physical Review B (Condensed Matter and Materials Physics)},
  \textbf{75}(2), 024\,515--8.

\bibitem[{Padilla \emph{et~al.}(2005)Padilla, Lee, Dumm, Blumberg, Ono, Segawa,
  Komiya, Ando \& Basov}]{delta1}
Padilla, W.~J., Lee, Y.~S., Dumm, M., Blumberg, G., Ono, S., Segawa, K.,
  Komiya, S., Ando, Y. \& Basov, D.~N. 2005 Constant effective mass across the
  phase diagram of high- $tc$ cuprates.
\newblock \emph{Phys. Rev. B}, \textbf{72}(6), 060\,511.
\newblock (\doi{10.1103/PhysRevB.72.060511})

\bibitem[{Park \emph{et~al.}(2008)Park, Haule \& Kotliar}]{imada}
Park, H., Haule, K. \& Kotliar, G. 2008 Cluster dynamical mean field theory of
  the mott transition.
\newblock \emph{Physical Review Letters}, \textbf{101}(18), 186\,403--4.

\bibitem[{Pasupathy \emph{et~al.}(2008)Pasupathy, Pushp, Gomes, Parker, Wen,
  Xu, Gu, Ono, Ando \emph{et~al.}}]{stripes5}
Pasupathy, A.~N., Pushp, A., Gomes, K.~K., Parker, C.~V., Wen, J., Xu, Z., Gu,
  G., Ono, S., Ando, Y. \emph{et~al.} 2008 Electronic origin of the
  inhomogeneous pairing interaction in the high-tc superconductor
  bi2sr2cacu2o8+{delta}.
\newblock \emph{Science}, \textbf{320}(5873), 196--201.

\bibitem[{Peets \emph{et~al.}(2009)Peets, Hawthorn, Shen, Kim, Ellis, Zhang,
  Komiya, Ando, Sawatzky \emph{et~al.}}]{peets}
Peets, D.~C., Hawthorn, D.~G., Shen, K.~M., Kim, Y.-J., Ellis, D.~S., Zhang,
  H., Komiya, S., Ando, Y., Sawatzky, G.~A. \emph{et~al.} 2009 X-ray absorption
  spectra reveal the inapplicability of the single-band hubbard model to
  overdoped cuprate superconductors.
\newblock \emph{Physical Review Letters}, \textbf{103}(8), 087\,402--4.

\bibitem[{Pellegrin \emph{et~al.}(1993)Pellegrin, N{\"u}cker, Fink, Molodtsov,
  Guti{\'e}rrez, Navas, Strebel, Hu, Domke \emph{et~al.}}]{pellegrin}
Pellegrin, E., N{\"u}cker, N., Fink, J., Molodtsov, S.~L., Guti{\'e}rrez, A.,
  Navas, E., Strebel, O., Hu, Z., Domke, M. \emph{et~al.} 1993 Orbital
  character of states at the fermi level in la2-xsrxcuo4 and r2-xcexcuo4
  (r=nd,sm).
\newblock \emph{Physical Review B}, \textbf{47}(6).

\bibitem[{Pereg-Barnea \emph{et~al.}(2010)Pereg-Barnea, Weber, Refael \&
  Franz}]{barnea}
Pereg-Barnea, T., Weber, H., Refael, G. \& Franz, M. 2010 Quantum oscillations
  from fermi arcs.
\newblock \emph{Nat Phys}, \textbf{6}(1), 44--49.

\bibitem[{Phillips(2010)}]{mottness}
Phillips, P. 2010 Colloquium: Identifying the propagating charge modes in doped
  mott insulators.
\newblock \emph{Rev. Mod. Phys.}, \textbf{82}(2), 1719--1742.
\newblock (\doi{10.1103/RevModPhys.82.1719})

\bibitem[{Phillips \& Chamon(2005)}]{pchamon}
Phillips, P. \& Chamon, C. 2005 Breakdown of one-parameter scaling in quantum
  critical scenarios for high-temperature copper-oxide superconductors.
\newblock \emph{Phys. Rev. Lett.}, \textbf{95}(10), 107\,002.
\newblock (\doi{10.1103/PhysRevLett.95.107002})

\bibitem[{Phillips \emph{et~al.}(2009{\natexlab{\emph{a}}})Phillips, Choy \&
  Leigh}]{charge2e3}
Phillips, P., Choy, T.-P. \& Leigh, R.~G. 2009{\natexlab{\emph{a}}} Mottness in
  high-temperature copper-oxide superconductors.
\newblock \emph{Reports on Progress in Physics}, \textbf{72}(3), 036\,501
  (24pp).

\bibitem[{Phillips \emph{et~al.}(2009{\natexlab{\emph{b}}})Phillips, Choy \&
  Leigh}]{repprogphys}
Phillips, P., Choy, T.-P. \& Leigh, R.~G. 2009{\natexlab{\emph{b}}} Mottness in
  high-temperature copper-oxide superconductors.
\newblock \emph{Reports on Progress in Physics}, \textbf{72}(3), 036\,501
  (24pp).

\bibitem[{Polchinski(1992)}]{polchinski}
Polchinski, J. 1992 Effective field theory and the fermi surface.

\bibitem[{Presland \emph{et~al.}(1991)Presland, Tallon, Buckley, Liu \&
  Flower}]{preslund}
Presland, M.~R., Tallon, J.~L., Buckley, R.~G., Liu, R.~S. \& Flower, N.~E.
  1991 General trends in oxygen stoichiometry effects on tc in bi and tl
  superconductors.
\newblock \emph{Physica C: Superconductivity}, \textbf{176}(1-3), 95--105.

\bibitem[{Randeria \emph{et~al.}(1992)Randeria, Trivedi, Moreo \&
  Scalettar}]{inco1}
Randeria, M., Trivedi, N., Moreo, A. \& Scalettar, R.~T. 1992 Pairing and spin
  gap in the normal state of short coherence length superconductors.
\newblock \emph{Phys. Rev. Lett.}, \textbf{69}(13), 2001--2004.
\newblock (\doi{10.1103/PhysRevLett.69.2001})

\bibitem[{Ranninger \emph{et~al.}(1995)Ranninger, Robin \& Eschrig}]{inco2}
Ranninger, J., Robin, J.~M. \& Eschrig, M. 1995 Superfluid precursor effects in
  a model of hybridized bosons and fermions.
\newblock \emph{Phys. Rev. Lett.}, \textbf{74}(20), 4027--4030.
\newblock (\doi{10.1103/PhysRevLett.74.4027})

\bibitem[{Rosch(2007)}]{zeros3}
Rosch, A. 2007 Breakdown of luttinger's theorem in two-orbital mott insulators.
\newblock \emph{The European Physical Journal B}, \textbf{59}(4), 495--502.
\newblock (\doi{10.1140/epjb/e2007-00312-3})

\bibitem[{Senthil \& Lee(2009)}]{senthil}
Senthil, T. \& Lee, P.~A. 2009 Synthesis of the phenomenology of the underdoped
  cuprates.
\newblock \emph{Physical Review B (Condensed Matter and Materials Physics)},
  \textbf{79}(24), 245116.
\newblock (\doi{10.1103/PhysRevB.79.245116})

\bibitem[{Shankar(1994)}]{shankar}
Shankar, R. 1994 Renormalization-group approach to interacting fermions.
\newblock \emph{Rev. Mod. Phys.}, \textbf{66}(1), 129--192.
\newblock (\doi{10.1103/RevModPhys.66.129})

\bibitem[{Shastry \emph{et~al.}(1993)Shastry, Shraiman \& Singh}]{shastryhall}
Shastry, B.~S., Shraiman, B.~I. \& Singh, R. R.~P. 1993 Faraday rotation and
  the hall constant in strongly correlated fermi systems.
\newblock \emph{Phys. Rev. Lett.}, \textbf{70}(13), 2004--2007.
\newblock (\doi{10.1103/PhysRevLett.70.2004})

\bibitem[{Simon \& Varma(2002)}]{trsb2}
Simon, M.~E. \& Varma, C.~M. 2002 Detection and implications of a time-reversal
  breaking state in underdoped cuprates.
\newblock \emph{Phys. Rev. Lett.}, \textbf{89}(24), 247\,003.
\newblock (\doi{10.1103/PhysRevLett.89.247003})

\bibitem[{Stanescu \& Kotliar(2006)}]{kotliar}
Stanescu, T.~D. \& Kotliar, G. 2006 Fermi arcs and hidden zeros of the green
  function in the pseudogap state.
\newblock \emph{Physical Review B (Condensed Matter and Materials Physics)},
  \textbf{74}(12), 125110.
\newblock (\doi{10.1103/PhysRevB.74.125110})

\bibitem[{Stanescu \& Phillips(2004)}]{stanescuhall}
Stanescu, T.~D. \& Phillips, P. 2004 Nonperturbative approach to full mott
  behavior.
\newblock \emph{Phys. Rev. B}, \textbf{69}(24), 245\,104.
\newblock (\doi{10.1103/PhysRevB.69.245104})

\bibitem[{Taillefer(2009)}]{tailifer}
Taillefer 2009 Fermi surface reconstruction in high-tc superconductors.
\newblock \emph{arXiv:0901.2313}.

\bibitem[{Takagi \emph{et~al.}(1989)Takagi, Ido, Ishibashi, Uota, Uchida \&
  Tokura}]{hall}
Takagi, H., Ido, T., Ishibashi, S., Uota, M., Uchida, S. \& Tokura, Y. 1989
  Superconductor-to-nonsuperconductor transition in ($la1-x$$srx$$)2$$cuo4$ as
  investigated by transport and magnetic measurements.
\newblock \emph{Phys. Rev. B}, \textbf{40}(4), 2254--2261.
\newblock (\doi{10.1103/PhysRevB.40.2254})

\bibitem[{Timusk \& Statt(1999)}]{timusk}
Timusk, T. \& Statt, B. 1999 The pseudogap in high-temperature superconductors:
  an experimental survey.
\newblock \emph{Reports on Progress in Physics}, \textbf{62}(1), 61--122.

\bibitem[{Tokura \emph{et~al.}(1988)Tokura, Torrance, Huang \& Nazzal}]{tokura}
Tokura, Y., Torrance, J.~B., Huang, T.~C. \& Nazzal, A.~I. 1988 Broader
  perspective on the high-temperature superconducting yba2cu3oy system: The
  real role of the oxygen content.
\newblock \emph{Physical Review B}, \textbf{38}(10).

\bibitem[{Tranquada \emph{et~al.}(2004)Tranquada, Woo, Perring, Goka, Gu, Xu,
  Fujita \& Yamada}]{stripes2}
Tranquada, J.~M., Woo, H., Perring, T.~G., Goka, H., Gu, G.~D., Xu, G., Fujita,
  M. \& Yamada, K. 2004 Quantum magnetic excitations from stripes in copper
  oxide superconductors.
\newblock \emph{Nature}, \textbf{429}(6991), 534--538.

\bibitem[{Tutsch \emph{et~al.}(1999)Tutsch, Schweiss, Hauff, Obst, Wolf \&
  Wahl}]{tutsch}
Tutsch, U., Schweiss, P., Hauff, R., Obst, B., Wolf, T. \& Wahl, H. 1999
  Calorimetric investigation of ndba2cu3ox single crystals.
\newblock \emph{Journal of Low Temperature Physics}, \textbf{117}(3), 951--955.

\bibitem[{Varma \emph{et~al.}(1989)Varma, Littlewood, Schmitt-Rink, Abrahams \&
  Ruckenstein}]{mfl}
Varma, C.~M., Littlewood, P.~B., Schmitt-Rink, S., Abrahams, E. \& Ruckenstein,
  A.~E. 1989 Phenomenology of the normal state of cu-o high-temperature
  superconductors.
\newblock \emph{Physical Review Letters}, \textbf{63}(18).

\bibitem[{Xia \emph{et~al.}(2008)Xia, Schemm, Deutscher, Kivelson, Bonn, Hardy,
  Liang, Siemons, Koster \emph{et~al.}}]{trsb1}
Xia, J., Schemm, E., Deutscher, G., Kivelson, S.~A., Bonn, D.~A., Hardy, W.~N.,
  Liang, R., Siemons, W., Koster, G. \emph{et~al.} 2008 Polar kerr-effect
  measurements of the high-temperature yba[sub 2]cu[sub 3]o[sub 6 + x]
  superconductor: Evidence for broken symmetry near the pseudogap temperature.
\newblock \emph{Physical Review Letters}, \textbf{100}(12), 127002.
\newblock (\doi{10.1103/PhysRevLett.100.127002})

\bibitem[{Xu \emph{et~al.}(2000)Xu, Ong, Wang, Kakeshita \& Uchida}]{nernst}
Xu, Z.~A., Ong, N.~P., Wang, Y., Kakeshita, T. \& Uchida, S. 2000 Vortex-like
  excitations and the onset of superconducting phase fluctuation in underdoped
  la$_{2-x}$sr$_x$cuo$_4$.
\newblock \emph{Nature}, \textbf{406}(486-488).

\bibitem[{Yoshida \emph{et~al.}(2006{\natexlab{\emph{a}}})Yoshida, Zhou,
  Tanaka, Yang, Hussain, Shen, Fujimori, Sahrakorpi, Lindroos
  \emph{et~al.}}]{shenweakcoupling}
Yoshida, T., Zhou, X.~J., Tanaka, K., Yang, W.~L., Hussain, Z., Shen, Z.-X.,
  Fujimori, A., Sahrakorpi, S., Lindroos, M. \emph{et~al.}
  2006{\natexlab{\emph{a}}} Systematic doping evolution of the underlying fermi
  surface of $ la2-{}x srx cu o4 $.
\newblock \emph{Phys. Rev. B}, \textbf{74}(22), 224\,510.
\newblock (\doi{10.1103/PhysRevB.74.224510})

\bibitem[{Yoshida \emph{et~al.}(2006{\natexlab{\emph{b}}})Yoshida, Zhou,
  Tanaka, Yang, Hussain, Shen, Fujimori, Sahrakorpi, Lindroos
  \emph{et~al.}}]{srd214b}
Yoshida, T., Zhou, X.~J., Tanaka, K., Yang, W.~L., Hussain, Z., Shen, Z.-X.,
  Fujimori, A., Sahrakorpi, S., Lindroos, M. \emph{et~al.}
  2006{\natexlab{\emph{b}}} Systematic doping evolution of the underlying fermi
  surface of la[sub 2 - x]sr[sub x]cuo[sub 4].
\newblock \emph{Physical Review B (Condensed Matter and Materials Physics)},
  \textbf{74}(22), 224510.
\newblock (\doi{10.1103/PhysRevB.74.224510})

\bibitem[{Yoshizaki \emph{et~al.}(1990)Yoshizaki, Ishikawa, Sawada, Kita \&
  Tasaki}]{yoshizaki}
Yoshizaki, R., Ishikawa, N., Sawada, H., Kita, E. \& Tasaki, A. 1990 Magnetic
  susceptibility of normal state and superconductivity of la2-xsrxcuo4.
\newblock \emph{Physica C: Superconductivity}, \textbf{166}(5-6), 417 -- 422.
\newblock (\doi{DOI: 10.1016/0921-4534(90)90038-G})

\bibitem[{Zaanen \& Gunnarsson(1989)}]{stripes3}
Zaanen, J. \& Gunnarsson, O. 1989 Charged magnetic domain lines and the
  magnetism of high-$tc$ oxides.
\newblock \emph{Phys. Rev. B}, \textbf{40}(10), 7391--7394.
\newblock (\doi{10.1103/PhysRevB.40.7391})

\end{thebibliography}
\end{document}